\documentclass[transmag]{IEEEtran}
\usepackage{latexsym}
\usepackage{graphicx}
\usepackage{amsfonts,amssymb,amsmath}
\usepackage{cite}
\usepackage{multirow}
\usepackage{graphicx}
\usepackage{gensymb}
\usepackage{xurl}

\usepackage{xcolor}

\usepackage{caption}
\usepackage{subcaption}

\def\BibTeX{{\rm B\kern-.05em{\sc i\kern-.025em b}\kern-.08em T\kern-.1667em\lower.7ex\hbox{E}\kern-.125emX}}
\begin{document}

\title{Temporary Laser Inter-Satellite Links in Free-Space Optical Satellite Networks}

\author{Aizaz U. Chaudhry, \IEEEmembership{Senior Member, IEEE}, and Halim Yanikomeroglu,
\IEEEmembership{Fellow, IEEE}
\thanks{Manuscript received July 26, 2022; revised August 8, 2022.
	
This work was supported by the High Throughput and Secure Networks Challenge Program at the National Research Council of Canada.}

\thanks{Aizaz U. Chaudhry and Halim Yanikomeroglu are with the Department of Systems and Computer Engineering, Carleton University, Ottawa, ON K1S 5B6, Canada (email: auhchaud@sce.carleton.ca; halim@sce.carleton.ca).}}

\IEEEtitleabstractindextext{\begin{abstract}\emph{Laser inter-satellite links} (LISLs) between satellites in a \emph{free-space optical satellite network} (FSOSN) can be divided into two classes: \emph{permanent LISLs} (PLs) and \emph{temporary LISLs} (TLs). TLs are not desirable in \emph{next-generation FSOSNs} (NG-FSOSNs) due to high LISL setup time, but they may become feasible in \emph{next-next-generation FSOSNs} (NNG-FSOSNs). Using the satellite constellation for Phase I of Starlink, we study the impact of TLs on network latency in an NG-FSOSN (which has only PLs) versus an NNG-FSOSN (which has PLs and TLs) under different long-distance inter-continental data communications scenarios, including Sydney–Sao Paulo, Toronto–Istanbul, Madrid–Tokyo, and New York–Jakarta, and different LISL ranges for satellites, including 659.5 km, 1,319 km, 1,500 km, 1,700 km, 2,500 km, 3,500 km, and 5,016 km. It is observed from the results that TLs provide higher satellite connectivity and thereby higher network connectivity, and they lead to lower average network latency for the NNG-FSOSN compared to the NG-FSOSN in all scenarios at all LISL ranges. In comparison with the NG-FSOSN, the improvement in latency with the NNG-FSOSN is significant at LISL ranges of 1,500 km, 1,700 km, and 2,500 km, where the improvement is 16.83 ms, 23.43 ms, and 18.20 ms, respectively, for the Sydney--Sao Paulo inter-continental connection. For the Toronto--Istanbul, Madrid--Tokyo, and New York--Jakarta inter-continental connections, the improvement is 14.58 ms, 23.35 ms, and 23.52 ms, respectively, at the 1,700 km LISL range.\end{abstract}


}

\maketitle

\section{INTRODUCTION}

\IEEEPARstart{S}{pacex} \cite{b1,b1a,b1b} and Telesat \cite{b2},\cite{b2a} are among several companies who plan to deploy their own low Earth orbit (LEO) and/or very low Earth orbit (VLEO) satellite constellations. The satellite constellation for Phase I of SpaceX’s Starlink is currently being deployed, and SpaceX has already started offering test satellite Internet services to individual users \cite{b3} while Telesat’s Lightspeed will start offering services in 2023 \cite{b4}. Unlike SpaceX’s Starlink, Telesat’s Lightspeed will provide business-to-business broadband Internet connectivity, whereas consumer broadband Internet is not their focus.

\par \emph{Laser inter-satellite links} (LISLs) are deemed essential in creating \emph{free-space optical satellite networks} (FSOSNs) \cite{b5}. Both SpaceX and Telesat are looking to equip their Starlink and Lightspeed satellites with LISLs \cite{b4},\cite{b6}. Starlink satellites launched on January 24, 2021, were equipped with \emph{laser communication terminals} (LCTs) for establishing LISLs. Only Starlink satellites in polar orbits had LCTs in 2021, whereas all Starlink satellites launched in 2022 are expected to be equipped with LCTs to create LISLs and establish an FSOSN.

\par LCTs for creating LISLs are still in their infancy. Only a few companies, like Mynaric and Tesat, are developing LCTs for satellites in upcoming LEO/VLEO constellations. Mynaric’s LCT for LEO satellites, CONDOR, is expected to offer LISLs between satellites with capacities of 10 Gbps at an LISL range of 4,500 km \cite{b7}. ConLCT1550 is the LCT being developed by Tesat specifically for LEO satellites, and it is promising to provide LISLs with 10 Gbps capacity at an LISL range of 6,000 km\cite{b8}. The LISL range of a satellite is a range over which it can establish an LISL with any other satellite that is within this range.

\par LISLs have been classified into two main categories \cite{b9}. The classification of LISLs in the first category is based on the location of satellites within a constellation. In this category, the LISLs between satellites are divided into four types: \emph{intra-orbital plane LISLs} between satellites in the same orbital plane (OP) (see Fig. 1); \emph{adjacent OP LISLs} between satellites in adjacent (left or right) OPs (see Fig. 2); \emph{nearby OP LISLs} between satellites in nearby OPs (see Fig. 3); and \emph{crossing OP LISLs} between satellites in crossing OPs (see Fig. 4). Nearby OPs for a satellite lie after adjacent OPs, and satellites in these OPs move in a direction similar to that of the satellite.       

\par In the second category, LISLs are classified based on the duration of their existence between satellites, and are divided into two types: \emph{permanent LISLs} (PLs) between a satellite and its neighbors that are always within its LISL range; and \emph{temporary LISLs} (TLs) between a satellite and other satellites that temporarily come within its range for a short period during their orbits. Intra-OP LISLs with satellites in the same OP are permanent in nature, and adjacent OP LISLs and nearby OP LISLs are usually permanent, whereas crossing OP LISLs can only exist temporarily as they are formed between satellites in crossing OPs that are moving in different directions.

\par At lower latitudes, for example at the equator, there are fewer satellites within LISL range of a satellite while at higher latitudes, near the Polar regions, there are more. Hence, LISLs with some satellites in adjacent and nearby OPs can also exist temporarily at higher latitudes when they come within LISL range of a satellite at these latitudes. Such temporary adjacent OP LISLs and temporary nearby OP LISLs as well as crossing OP LISLs fall within the class of TLs. 

\par In this work, we investigate the impact of TLs on the latency of FSOSNs. In doing so, we employ the satellite constellation for Phase I of Starlink. We consider two types of FSOSNs: one having only PLs, while the other has PLs as well as TLs. We call the first type of FSOSN a \emph{next-generation FSOSN} (NG-FSOSN) and the second type a \emph{next-next-generation FSOSN} (NNG-FSOSN). As part of our study, we examine the connectivity among satellites in these networks at seven different LISL ranges, specifically 659.5 km, 1,319 km, 1,500 km, 1,700 km, 2,500 km, 3,500 km, and 5,016 km. Then, we compare the average network latency of these FSOSNs for long-distance inter-continental data communications between ground stations in cities over the shortest paths across these networks. For example, we compare the average network latency of their shortest paths at different LISL ranges for the Sydney--Sao Paulo inter-continental connection. Furthermore, we compare these FSOSNs under three other inter-continental connection scenarios, namely Toronto--Istanbul, Madrid--Tokyo, and New York--Jakarta, at two different LISL ranges: 1,700 km (a low LISL range for satellites in Starlink's Phase I constellation) and 5,016 km (the highest LISL range for satellites in this constellation).

\par We observe that satellite connectivity with TLs (i.e., with PLs and TLs) is at least twice that without TLs (i.e., with only PLs). The satellite connectivity varies with latitudes and is higher at higher latitudes. The network connectivity in the NNG-FSOSN (i.e., the FSOSN with PLs and TLs) is also at least twice that in the NG-FSOSN (i.e., the FSOSN with only PLs). For the Sydney--Sao Paulo inter-continental connection scenario, a comparison of the two FSOSNs in terms of average network latency could not be made at lower LISL ranges of 659.5 km and 1,319 km due to the unavailability of shortest paths across the NG-FSOSN at these ranges. For higher LISL ranges of 1,500 km or more, it is observed that the average network latency of the NNG-FSOSN (with TLs) is consistently better than that of the NG-FSOSN (without TLs). A similar trend is observed for the Toronto--Istanbul, Madrid--Tokyo, and New York--Jakarta inter-continental connection scenarios at the 1,700 km and 5,016 km LISL ranges. To the best of our knowledge, this work is the first to study the effect of TLs on network latency (or more precisely average network latency) of an NG-FSOSN versus an NNG-FSOSN. 

\par The rest of the paper is organized as follows. A brief discussion of NG- and NNG-FSOSNs is provided in Section II along with the related work and the motivation for our work. Section III examines the impact of TLs on satellite connectivity and network connectivity. The effect of TLs on network latency of the two types of FSOSNs is investigated in Section IV, and some design guidelines are provided. Section V summarizes our findings and offers directions for future work.

\begin{figure}[htbp]
	\centerline{\includegraphics[scale=0.39]{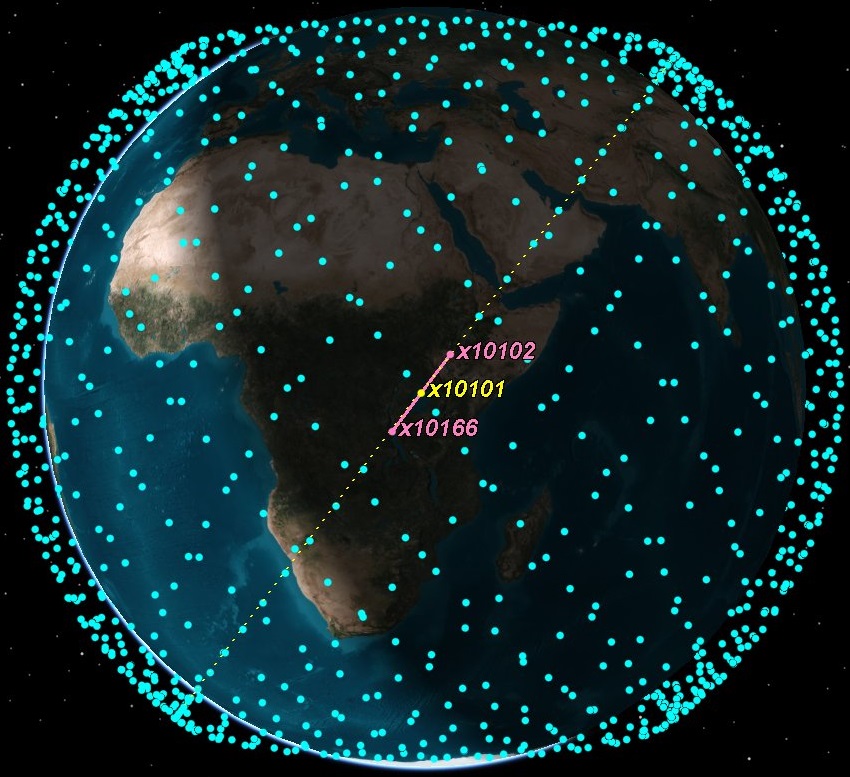}}
	\renewcommand\thefigure{1}\caption{An example of intra-orbital plane LISLs between satellite \emph{x10101} (i.e., the first satellite in the first orbital plane of Starlink's Phase I constellation) and satellites \emph{x10102} and \emph{x10166} is shown in this figure. The LISLs are shown by solid pink lines, while the orbital plane of satellites \emph{x10101}, \emph{x10102}, and \emph{x10166} is shown by the dashed yellow line.}
\end{figure}

\begin{figure}[htbp]
	\centerline{\includegraphics[scale=0.39]{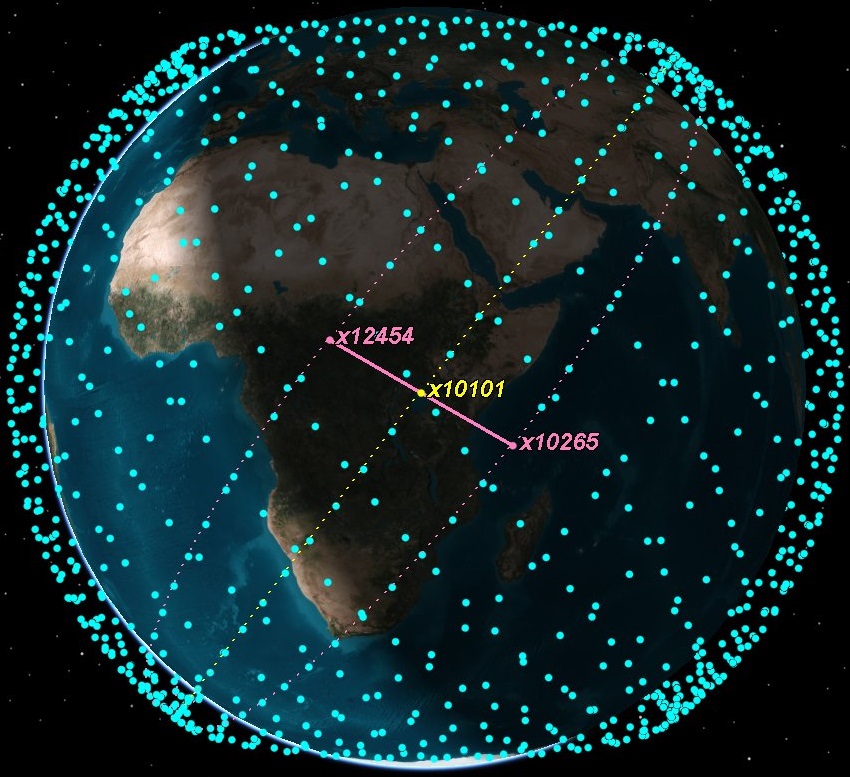}}
	\renewcommand\thefigure{2}\caption{An example of adjacent orbital plane LISLs between satellite \emph{x10101} and satellites \emph{x10265} and \emph{x12454} is shown in this figure. The LISLs are shown by solid pink lines, the orbital plane of satellite \emph{x10101} is shown by the dashed yellow line, while the orbital planes of satellites \emph{x10265} and \emph{x12454} are shown by dashed pink lines.}
\end{figure}

\begin{figure}[htbp]
	\centerline{\includegraphics[scale=0.39]{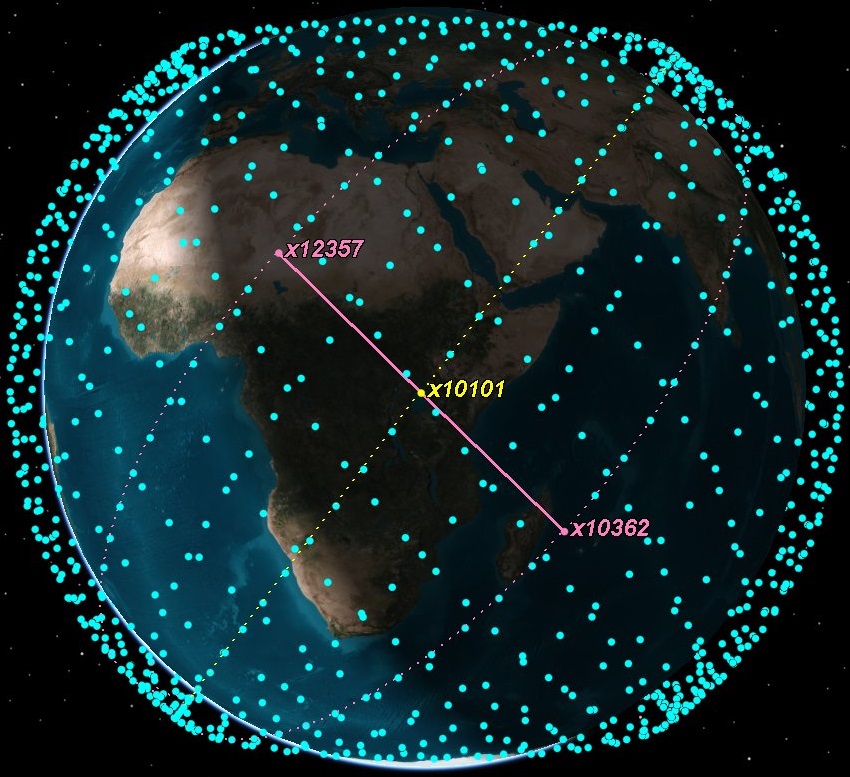}}
	\renewcommand\thefigure{3}\caption{An example of nearby orbital plane LISLs between satellite \emph{x10101} and satellites \emph{x10362} and \emph{x12357} is shown in this figure. The LISLs are shown by solid pink lines, the orbital plane of satellite \emph{x10101} is shown by the dashed yellow line, while the orbital planes of satellites \emph{x10362} and \emph{x12357} are shown by dashed pink lines.}
\end{figure}

\begin{figure}[htbp]
	\centerline{\includegraphics[scale=0.39]{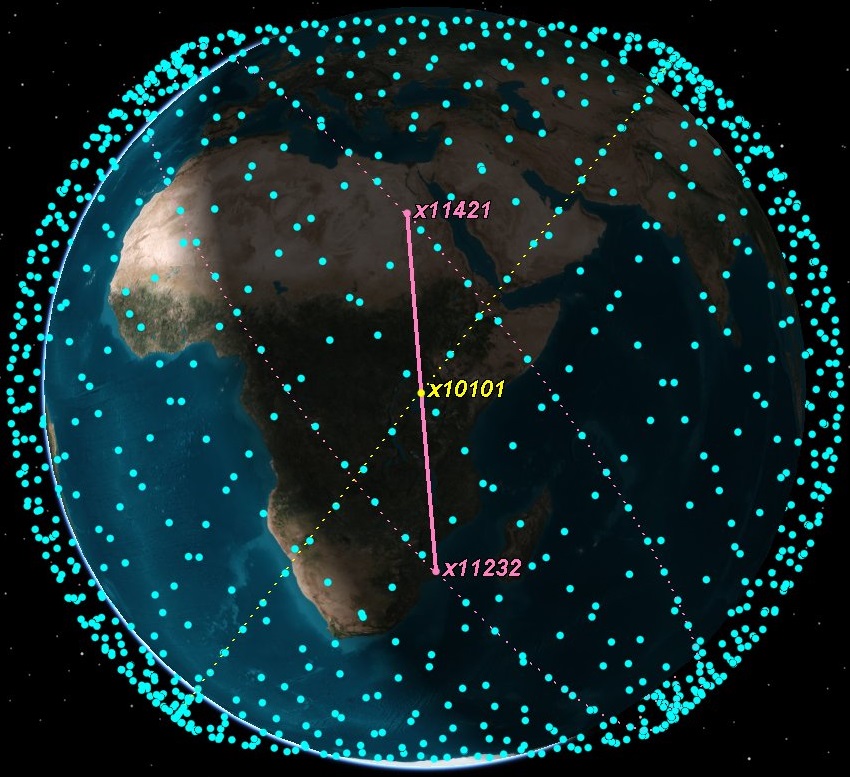}}
	\renewcommand\thefigure{4}\caption{An example of crossing orbital plane LISLs between satellite \emph{x10101} and satellites \emph{x11232} and \emph{x11421} is shown in this figure. The LISLs are shown by solid pink lines, the orbital plane of satellite \emph{x10101} is shown by the dashed yellow line, while the orbital planes of satellites \emph{x11232} and \emph{x11421} are shown by dashed pink lines.}
\end{figure}

\section{Motivation and Related Work}

\par FSOSNs are emerging as a promising solution to provide low-latency long-distance inter-continental data communications between financial stock markets \cite{b9a}, where a 1 millisecond advantage in latency in high-frequency trading can save \$100 million a year for a single brokerage firm \cite{b9b}, and a reduction in latency of a few milliseconds can translate into billions of dollars of revenues for such firms. An FSOSN can offer low-latency communications as a premium service to the financial centers around the globe and this use-case can help in recovering the deployment and maintenance cost of FSOSNs.

\par The current setup times to establish LISLs between satellites range from a few seconds to tens of seconds due to the pointing, acquisition, and tracking (PAT) process \cite{b9c} involved at LCTs in creating LISLs. For example, Mynaric’s CONDOR LCT requires an initial acquisition time of around 30 seconds. However, once the network is created and the position and altitude information of satellites is exchanged among all satellites, the acquisition time for this LCT reduces to two seconds \cite{b10}. Due to these high LISL setup times, TLs are currently considered undesirable.

\par The NG-FSOSNs created by employing LISLs between satellites in upcoming LEO/VLEO constellations, like Starlink and Lightspeed, are expected to become fully operational by the mid- to late-2020s. The high LISL setup times offered by existing LCTs will restrict satellites in these NG-FSOSNs to only establishing PLs with other satellites \cite{b10aa}. Once the PLs are established between satellites, and the satellite network is created, these LISLs will need to operate continuously. However, creating new LISLs and avoiding corresponding LISL setup delays may not be avertable. Satellites in these NG-FSOSNs may fail creating the need to set up new PLs between satellites to route around the failed satellite. 

 \par In addition to the prohibitive LISL setup times, the complexity in establishing TLs with temporary neighbors---such as crossing OP neighbors (which temporarily come within range while crossing) and temporary adjacent OP and temporary nearby OP neighbors (which  temporarily come within range at high latitudes)---moving at high relative velocities requires a highly sophisticated and accurate PAT system to establish and maintain such TLs. The creation of an LISL between a pair of satellites involves pointing a laser beam at a moving satellite from another moving satellite \cite{b10a1}. This is a bigger challenge in establishing TLs due to the higher relative velocities of satellites. Per their 2016 FCC filing, SpaceX initially planned to equip its Starlink satellites with five LCTs each \cite{b1}. Four LCTs were to be used for establishing PLs with neighbors in the same orbital plane and in adjacent orbital planes, while the fifth LCT was intended to create a TL with a crossing OP neighbor. However, SpaceX reduced the number of LCTs per a Starlink satellite from five to four in a 2018 FCC filing \cite{b1b} due to the high complexity of developing an efficient PAT system suitable for TLs.

\par In NNG-FSOSNs (or third-generation FSOSNs), which are expected to follow second-generation FSOSNs (or NG-FSOSNs), beginning in the early- to mid-2030s, LISL setup times are likely to be in milliseconds due to technological advancements. We envision that LCTs with such very low LISL setup times will become available in NNG-FSOSNs. Not only will such LCTs help in quickly circumventing satellite failure, they will also enable the use of TLs in addition to PLs, which will positively impact the latency performance of NNG-FSOSNs. The characteristics that differentiate NNG-FSOSNs from NG-FSOSNs are summarized in Table 1. The only operational example of a current-generation FSOSN (i.e., a first-generation FSOSN) is the European Data Relay System, which employs 1.8 Gbps LISLs between the system's geostationary Earth orbit satellites and client LEO satellites for data-relay services \cite{b10d},\cite{b10e}.

\begin{table*}
	\centering
	\renewcommand\thetable{1}\caption{Characteristics of NG-FSOSNs vs. NNG-FSOSNs.}
\begin{tabular}{|c|c|c|} 
	\hline
	\textbf{Characteristic} & \textbf{NG-FSOSNs}                                                                             & \textbf{NNG-FSOSNs}                                                                                                                                                                                                     \\ 
	\hline
	Type                               & \begin{tabular}[c]{@{}c@{}}Next-generation \\(or second-generation) FSOSNs\end{tabular}     & Next-next-generation (or third-generation) FSOSNs                                                                                                                                                                                             \\ 
	\hline
	Objective                          & \begin{tabular}[c]{@{}c@{}}Deliver broadband Internet to \\rural and remote areas \cite{b1b},\cite{b2a},\cite{b10a2}\end{tabular} & Deliver an all-optical transport network
	in space \cite{b10b,b10c,b10c1}                                                                                                                                                                     \\ 
	\hline
	Timeline for operational readiness & Mid- to late-2020s                                                                              & Early- to mid-2030s                                                                                                                                                                                                      \\ 
	\hline
	Data rate for LISLs                & Up to 10 Gbps                                                                                  & In Tbps                                                                                                                                                                                                                 \\ 
	\hline
	Capability of LCTs                 & \begin{tabular}[c]{@{}c@{}}Will support only \\permanent LISLs\end{tabular}                    & Will support permanent as well as temporary LISLs                                                                                                                                                                       \\ 
	\hline
	LISL setup time                    & 2 to 30 seconds                                                                                & In the order of a few milliseconds                                                                                                                                                                                                         \\ 
	\hline
	Cost of LCTs                       & High                                                                                           & Low                                                                                                                                                                                                                     \\ 
	\hline
	Number of LCTs per satellite       & \begin{tabular}[c]{@{}c@{}}Limited to four to support LISLs \\within an FSOSN\end{tabular}     & \begin{tabular}[c]{@{}c@{}}Five or more to support both intra-FSOSN LISLs \\(i.e., LISLs between satellites within an FSOSN), \\and inter-FSOSN LISLs (i.e., LISLs between satellites \\in different FSOSNs)\end{tabular}  \\ 
	\hline
	Standardization of LCTs            & ~
	Not standardized                                                                           & \begin{tabular}[c]{@{}c@{}}Standardized; \\satellites in different FSOSNs equipped with LCTs \\from different vendors will be able to establish LISLs \\and seamlessly communicate with each other\end{tabular}           \\
	\hline
\end{tabular}
\end{table*}

\par The study of latency in FSOSNs has attracted the attention of the research community \cite{b5},\cite{b9a,b10aa,b12,b13,b14,b15,b16a}. Typically made of glass, optical fibers have a refractive index of around 1.5; this corresponds to the speed of light in optical fiber, which is approximately 50\% lower than in the vacuum of space. This gives FSOSNs (also known as optical wireless satellite networks) a critical advantage over optical fiber terrestrial networks in terms of latency for long-distance inter-continental data communications. In \cite{b5}, a use-case was investigated to determine the suitability of optical wireless satellite networks for low-latency data communications over long distances in comparison with optical fiber terrestrial networks. It was noted that an optical wireless satellite network at 550 km altitude provided better latency than an optical fiber terrestrial network when the terrestrial hop distance was more than 3,000 km. 

\par A comparison of latency (in terms of propagation delay) for an optical wireless satellite network and an optical fiber terrestrial network in different scenarios for long-distance inter-continental data communications was provided in \cite{b9a}. LISLs were assumed between satellites in Starlink’s Phase I constellation to realize an FSOSN. It was observed that laser links between satellites and between satellites and ground stations and/or the latency of these links changes at every time slot due to the high orbital speed of satellites along their OPs, and thereby the shortest path between ground stations (in two different cities) over an FSOSN and/or its latency also changes at every time slot. The shortest distance between two cities in different continents over the optical fiber terrestrial network was used in this comparison; long-haul submarine optical fiber cables are not laid along the shortest path to connect two points on Earth’s surface; and thereby this comparison favored the optical fiber terrestrial network. Nevertheless, it was noted that the FSOSN provided lower latency than the optical fiber terrestrial network in all scenarios. 

\par The use of repetitive patterns in network topology, called motifs, was proposed in \cite{b10aa} to improve the latency of satellite networks. Within a motif, the same connectivity pattern was considered for all satellites within the satellite network. Different motifs represented different connectivity patterns; performance of different motifs was found to be totally different; and finding the motif with the best performance meant exhaustively assessing the performance of all motifs. A hypothetical constellation consisting of 1,600 satellites at an altitude of 550 km was used; a median \hbox{round-trip} time improvement of 70\% was observed as compared to Internet latency; however, delays, such as those arising from sub-optimal routing, congestion, queueing, and forward error correction were not considered in the satellite network but they were considered while measuring Internet latency.

\par A preliminary evaluation of how well a satellite network using LISLs can provide low-latency communications was conducted in \cite{b12}. Starlink’s original constellation for its Phase I was used, which comprised 1,600 satellites at an altitude of 1,150 km; LISLs were assumed between satellites in this constellation; and the latency of the resulting satellite network was investigated. In addition to four LISLs for each satellite with nearest neighbors in the same OP and in adjacent OPs, a nearby satellite in a crossing OP was also considered for establishing an LISL for better routing paths. It was concluded that this satellite network could offer lower latency communications than a terrestrial optical fiber network over longer distances. 

\par In \cite{b13}, it was observed that using LISLs between satellites provided lower latency than using ground-based relays. Starlink’s revised constellation for its Phase I was employed for this investigation, which comprised 1,584 satellites at an altitude of 550 km. Ground-based relays were considered as an alternative to LISLs to achieve low-latency wide area networking. It was mentioned that if an LISL could not provide sufficient capacity to meet the offered load, multipath routing could be used to divide the load and a ground-based relay could be employed to complement the LISL to supplement capacity in busy parts of the satellite network.    

\par Another study showed that employing LISLs between satellites in a constellation significantly reduced the temporal variations in latency of the resulting satellite network \cite{b14}. In this investigation, the satellite network realized through LISLs was compared with the bent-pipe scenario, where the connectivity between satellites in the constellation was achieved through ground stations. It was observed that LISLs could provide better throughput and lower attenuation than that achieved without such links in addition to substantially reducing the temporal variations in latency.

\par A simulator was developed in \cite{b15} to study the network behavior, including latency, of satellite networks arising from upcoming LEO/VLEO constellations. A packet-level simulation environment was provided by this simulator, which incorporated the orbital dynamics of a satellite network based on a LEO/VLEO constellation. Its visualization module could render views of satellite trajectories and ground station perspective on overhead satellites. This simulator was then used to analyze the behavior of connections, such as their latency, across satellite networks that were based on three upcoming satellite constellations. However, the connectivity of a satellite was limited to four nearest neighbors in this simulator and it did not support crossing OP or temporary LISLs. This simulator would have to be modified if it were to be used to investigate the performance of an FSOSN comprising PLs as well as TLs.

\par A crossover function was proposed in \cite{b16a}, which was then used to determine the crossover distance, where the crossover distance was defined as the distance between two points on the surface of the Earth beyond which crossing over from an optical fiber terrestrial network to an FSOSN could be useful for lower latency data communications. It was shown that the crossover distance depended on the refractive index of the optical fiber in the optical fiber terrestrial network and the altitude of satellites and the end-to-end propagation distance in the FSOSN. A TL with a nearby crossing OP neighbor was considered in \cite{b12} while studying latency of a satellite network, and TLs were also considered in \cite{b9a} and \cite{b16a} while conducting comparison of an optical wireless satellite network and an optical fiber terrestrial network in terms of latency. Yet no studies exist that examine an NG-FSOSN versus an NNG-FSOSN to investigate the impact of TLs on network latency of these FSOSNs.

\section{Impact of Temporary Laser Inter-Satellite Links on Satellite and Network Connectivity}

\par To study the impact of TLs on satellite connectivity and network connectivity, we consider the satellite constellation for Phase I of Starlink. It consists of 1,584 LEO satellites in 24 OPs with 66 satellites in each OP. The altitude of the satellites in this constellation is 550 km, their inclination is 53º with respect to the equator, and we assume this constellation to be uniform. We study the satellite connectivity and network connectivity at seven different LISL ranges for satellites varying from the minimum LISL range for this constellation to the maximum, i.e., from 659.5 km to 5,016 km.

\par The reasoning for choosing these seven LISL ranges to study the effect of TLs on satellite connectivity and network connectivity in an FSOSN based on Starlink’s Phase I constellation is discussed hereafter. An LISL range of 659.5 km is identified as the minimum LISL range for Starlink’s Phase I constellation, as a satellite at this range has minimum possible connectivity since it has PLs with only two intra-OP neighbors (see Fig. 5). A satellite in this constellation has PLs with four intra-OP neighbors at the 1,319 km LISL range (see Fig. 6); it has PLs with nearest two adjacent OP neighbors at the 1,500 km LISL range in addition to PLs with four intra-OP neighbors (see Fig. 7); and it has PLs with nearest six adjacent OP neighbors at the 1,700 km LISL range besides PLs with four intra-OP neighbors (see Fig. 8). The maximum possible LISL range for a satellite in this constellation is an LISL range that is limited only by visibility and it is calculated as 5,016 km \cite{b9}. Two intermediate LISL ranges of 2,500 km and 3,500 km are also considered for studying this effect at these ranges.

\begin{figure}[htbp]
	\centerline{\includegraphics[scale=0.445]{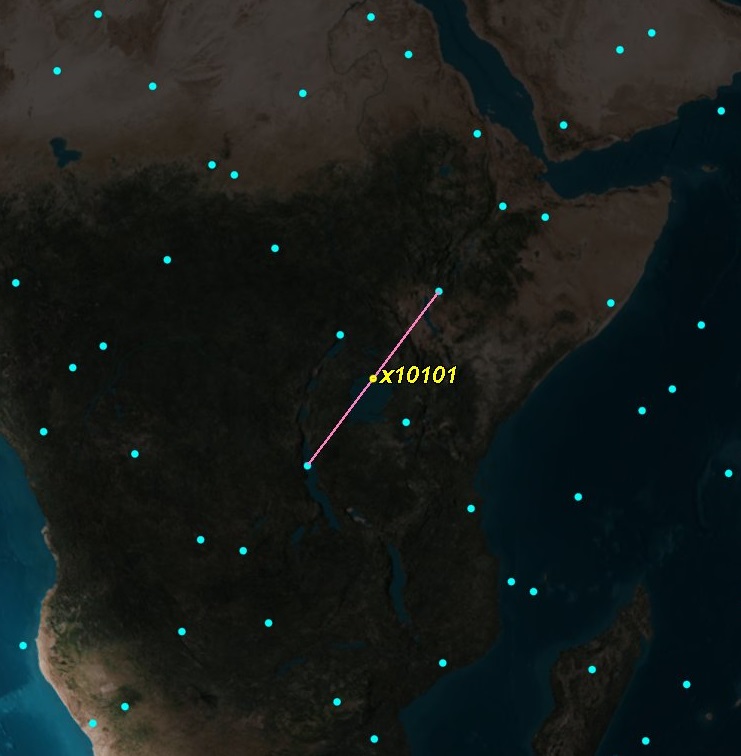}}
	\renewcommand\thefigure{5}\caption{The satellite connectivity for \emph{x10101} at the 659.5 km LISL range with only PLs at $0\degree$ latitude is shown in this figure. The number of possible PLs that this satellite can establish at this range is 2, as displayed in this figure.}
\end{figure}

\begin{figure}[htbp]
	\centerline{\includegraphics[scale=0.445]{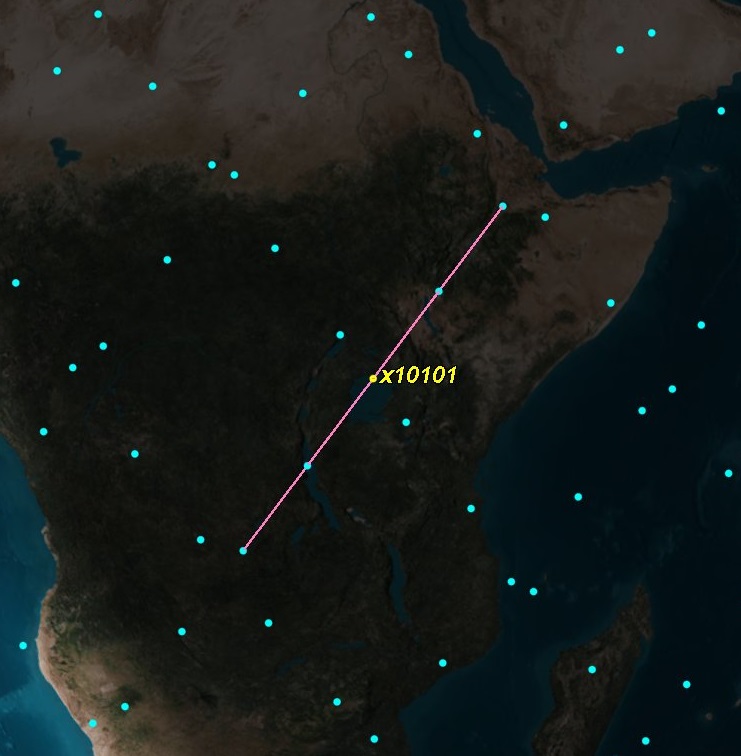}}
	\renewcommand\thefigure{6}\caption{The satellite connectivity for \emph{x10101} at the 1,319 km LISL range with only PLs at $0\degree$ latitude is shown in this figure. The number of possible PLs that this satellite can establish at this range is 4, as shown in this figure.}
\end{figure} 

\begin{figure}[htbp]
	\centerline{\includegraphics[scale=0.445]{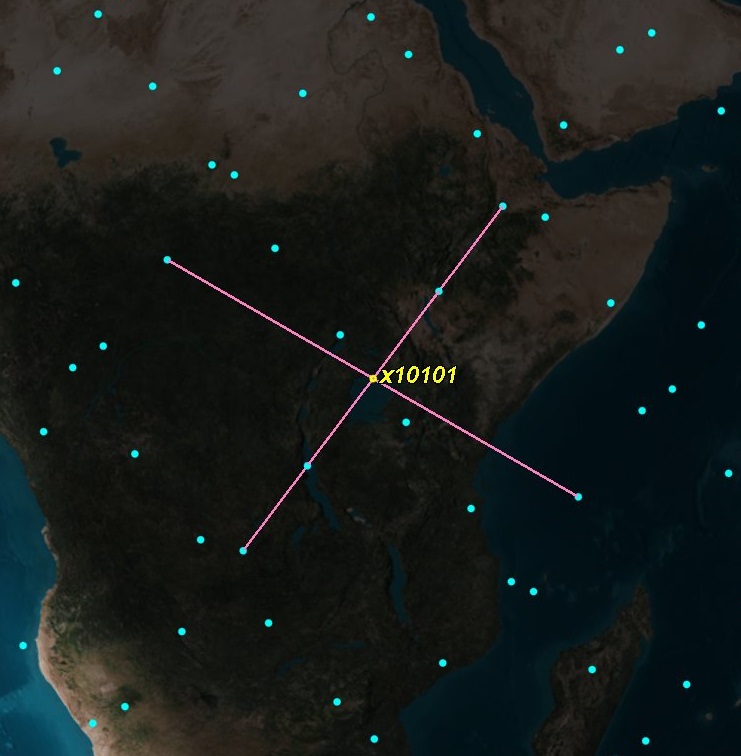}}
	\renewcommand\thefigure{7}\caption{The satellite connectivity for \emph{x10101} at the 1,500 km LISL range with only PLs at $0\degree$ latitude is shown in this figure. The number of possible PLs that this satellite can establish at this range is 6, as indicated in this figure.}
\end{figure}  

\par Table 2 shows a comparison of the satellite connectivity for satellite \emph{x10101} (i.e., the first satellite in the first OP of Starlink's Phase I constellation) for different LISL ranges with only PLs, with PLs and TLs at the equator at 0º latitude, and with PLs and TLs near the North Pole at 47.33º latitude. With only PLs, the connectivity (i.e., the number of permanent neighbors within the LISL range of the satellite or the number of possible permanent LISLs the satellite can establish with satellites that are permanently within its LISL range) of \emph{x10101} remains the same at the equator and near the Poles. For example, \emph{x10101} can form four possible PLs at the 1,319 km LISL range at all latitudes.

\begin{table*}
	\centering
	\renewcommand\thetable{2}\caption{Satellite Connectivity for \emph{x10101} with PLs vs. with PLs and TLs.}
	\begin{tabular}{|l|c|c|c|c|c|} 
		\hline
		\multirow{3}{*}{\begin{tabular}[c]{@{}l@{}}\textbf{LISL Range}\\\textbf{(km)}\end{tabular}} & \multicolumn{5}{c|}{\textbf{Number of possible LISLs with satellites within LISL range }}                                                                                                                                                                                                                             \\ 
		\cline{2-6}
		& \multirow{2}{*}{\textbf{PLs}} & \multirow{2}{*}{\begin{tabular}[c]{@{}c@{}}\textbf{PLs and TLs}\\\textbf{(at 0º latitude)}\end{tabular}} & \multirow{2}{*}{\begin{tabular}[c]{@{}c@{}}\textbf{PLs and TLs}\\\textbf{(at 47.33º latitude)}\end{tabular}} & \multicolumn{2}{c|}{\textbf{Improvement}}                \\ 
		\cline{5-6}
		&                                 &                                                                                                          &                                                                                                              & \textbf{At 0º latitude} & \textbf{At 47.33º latitude}  \\ 
		\hline
		659.5                                                                                          & 2                               & 4                                                                                                        & 8                                                                                                            & 2                         & 6                              \\ 
		\hline
		1,319                                                                                          & 4                               & 8                                                                                                        & 29                                                                                                           & 4                         & 25                             \\ 
		\hline
		1,500                                                                                          & 6                               & 12                                                                                                       & 33                                                                                                           & 6                         & 27                             \\ 
		\hline
		1,700                                                                                          & 10                              & 22                                                                                                       & 40                                                                                                           & 12                        & 30                             \\ 
		\hline
		2,500                                                                                          & 18                              & 38                                                                                                       & 70                                                                                                           & 20                        & 52                             \\ 
	    \hline
		3,500                                                                                          & 42                              & 88                                                                                                       & 117                                                                                                           & 46                        & 75                            \\ 
		\hline
		5,016                                                                                          & 88                              & 180                                                                                                      & 209                                                                                                          & 92                        & 121                            \\
		\hline
	\end{tabular}
\end{table*}

\par When we consider TLs in addition to PLs, \emph{x10101} has more temporary neighbors (i.e., temporary adjacent OP, temporary nearby OP, and crossing OP neighbors) within its range near the Poles and less at the equator due to the inclined nature of the constellation. For example, the number of possible LISLs for \emph{x10101} when considering PLs and TLs is 8 at 0º latitude and 29 at 47.33º latitude for the 1,319 km LISL range. At this range, all 4 TLs at 0º latitude are crossing OP LISLs while at 47.33º latitude, the 25 TLs consist of a mix of crossing OP LISLs, temporary adjacent OP LISLs, and temporary nearby OP LISLs.

\par It is also clear from Table 2 that the satellite connectivity for \emph{x10101} with PLs and TLs at any latitude is much higher than that with only PLs. It is at least twice with PLs and TLs compared to only PLs. For example, the number of possible PLs that \emph{x10101} can establish at the 1,700 km LISL range is 10 (see Fig. 8), the number of possible PLs and TLs at 0º latitude is 22 for \emph{x10101} (see Fig. 9), and the number of possible PLs and TLs at 47.33º latitude is 40 for this satellite (see Fig. 10).     

\begin{figure*}[htbp]
	\centerline{\includegraphics[scale=0.405]{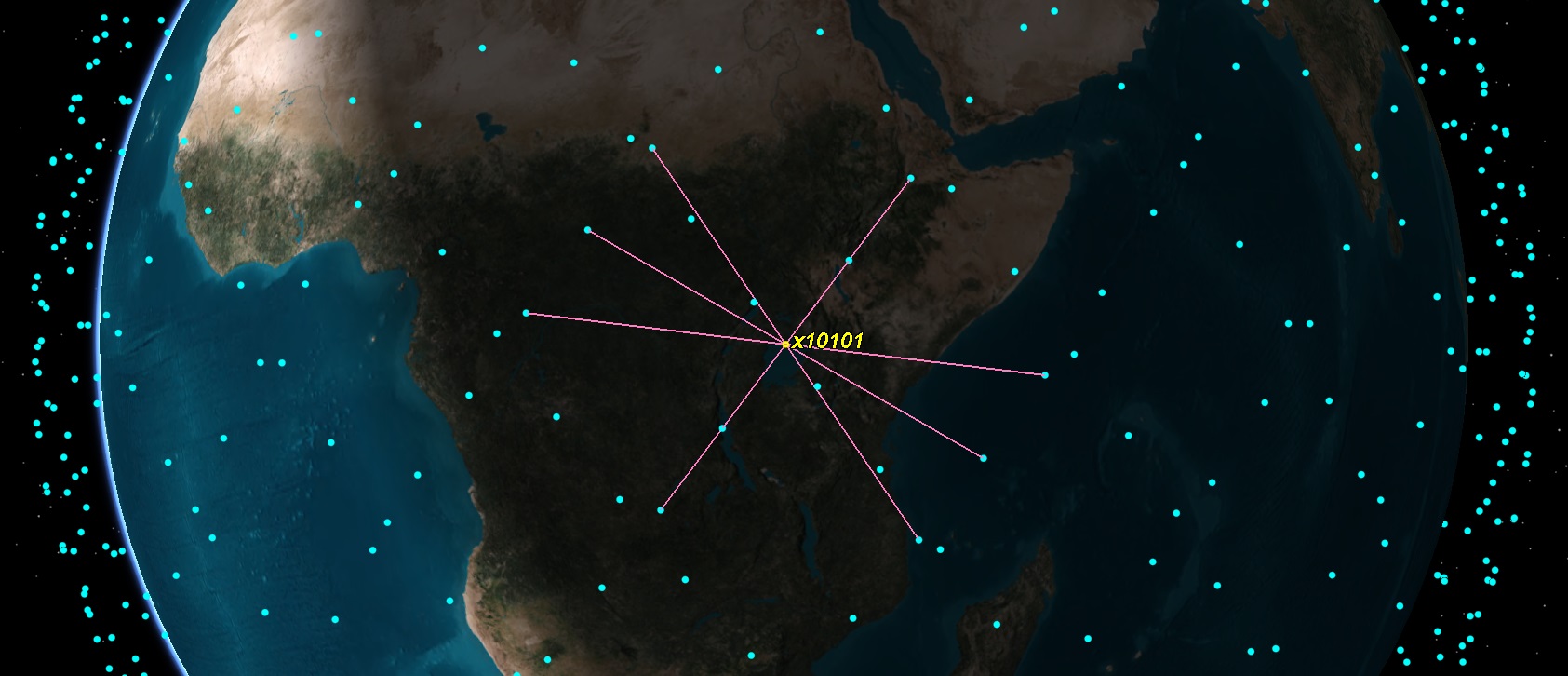}}
	\renewcommand\thefigure{8}\caption{The satellite connectivity for \emph{x10101} at the 1,700 km LISL range with only PLs at $0\degree$ latitude is shown in this figure. The number of possible PLs that this satellite can establish at this range is 10, as illustrated in this figure. With only PLs, the satellite connectivity remains the same at the equator as well as at higher latitudes near the Poles.}
\end{figure*}

\begin{figure*}[htbp]
	\centerline{\includegraphics[scale=0.405]{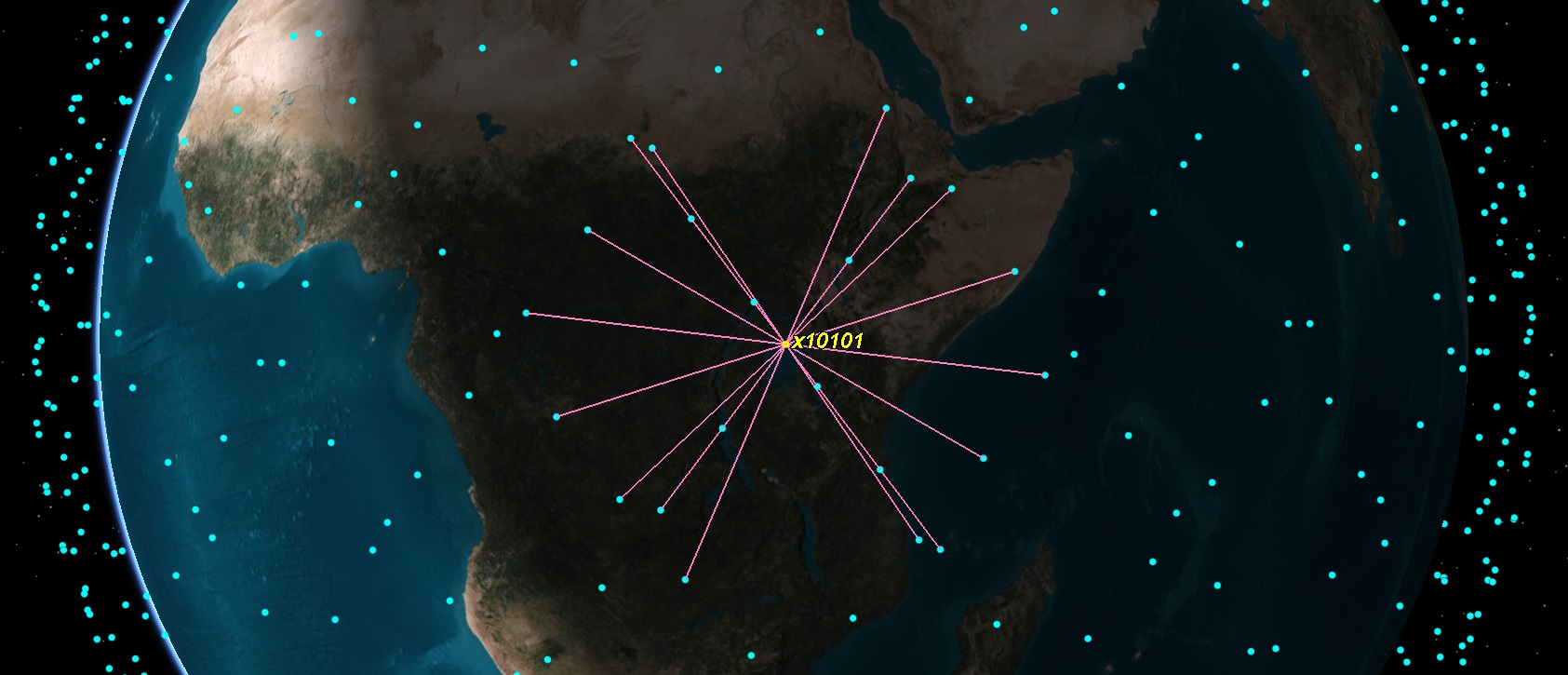}}
	\renewcommand\thefigure{9}\caption{The satellite connectivity for \emph{x10101} at the 1,700 km LISL range with PLs and TLs at $0\degree$ latitude is shown in this figure. The number of possible PLs and TLs that this satellite can establish at this latitude and range is 22, as highlighted in this figure. The satellite connectivity for \emph{x10101} with PLs and TLs at $0\degree$ latitude and 1,700 km LISL range is more than twice that with only PLs at this latitude and range.}
\end{figure*}

\begin{figure*}[htbp]
	\centerline{\includegraphics[scale=0.405]{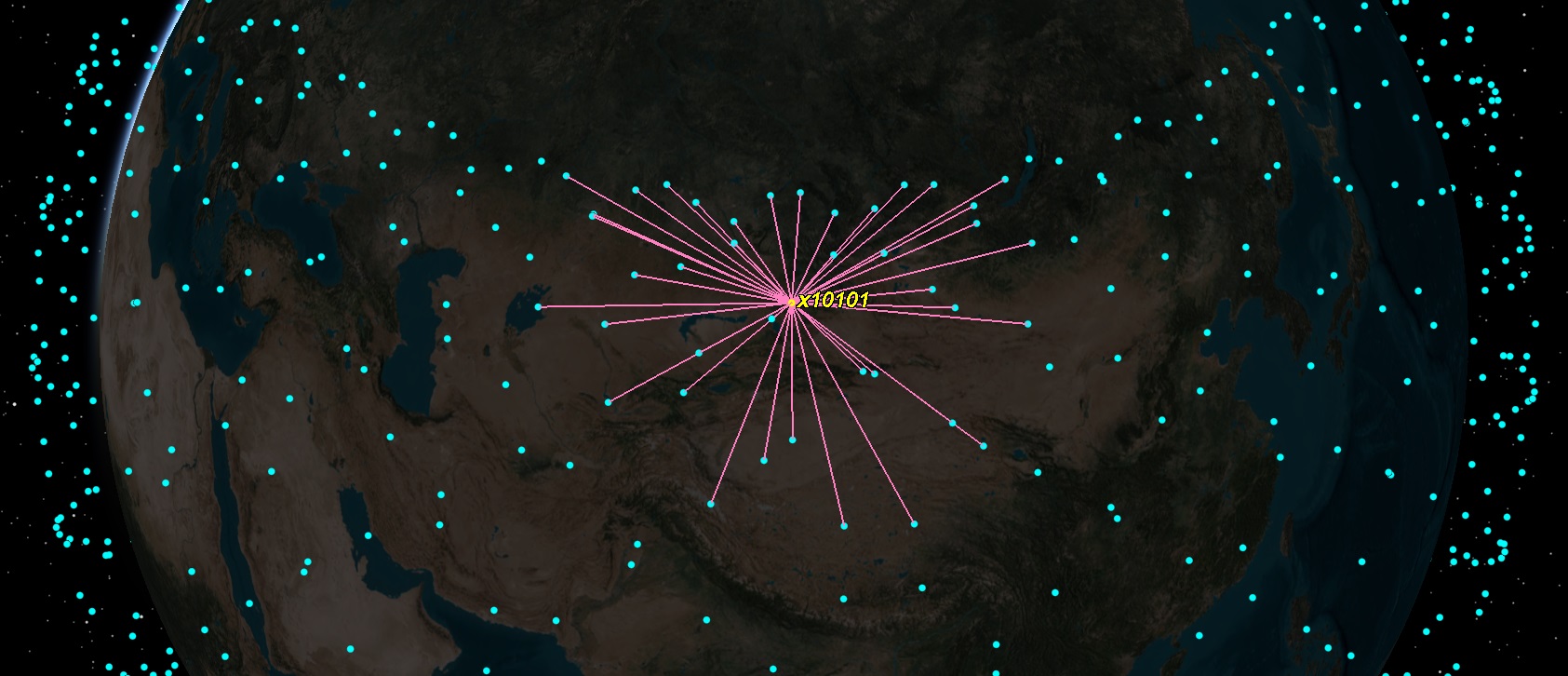}}
	\renewcommand\thefigure{10}\caption{The satellite connectivity for \emph{x10101} at the 1,700 km LISL range with PLs and TLs at $47.33\degree$ latitude is shown in this figure. The number of possible PLs and TLs that this satellite can establish at this latitude and range is 40, as displayed in this figure. The satellite connectivity for \emph{x10101} with PLs and TLs at $47.33\degree$ latitude and 1,700 km LISL range is four times that with only PLs at $0\degree$ latitude and this range, and it is approximately twice that with PLs and TLs at $0\degree$ latitude and this range. With PLs and TLs, higher satellite connectivity is achieved at higher latitudes near the Poles than at the equator.}
\end{figure*}

\par The connectivity of satellite \emph{x10101} increases in accordance with the increase in LISL range, as indicated by the results in Table 2. For example, \emph{x10101} has connectivity with 12 neighbors via PLs and TLs at the 1,500 km LISL range and 0º latitude, as illustrated in Fig. 11, and its connectivity using PLs and TLs increases to 38 at the 2,500 km LISL range and 0º latitude, as shown in Fig. 12.

\begin{figure}[htbp]
	\centerline{\includegraphics[scale=0.297]{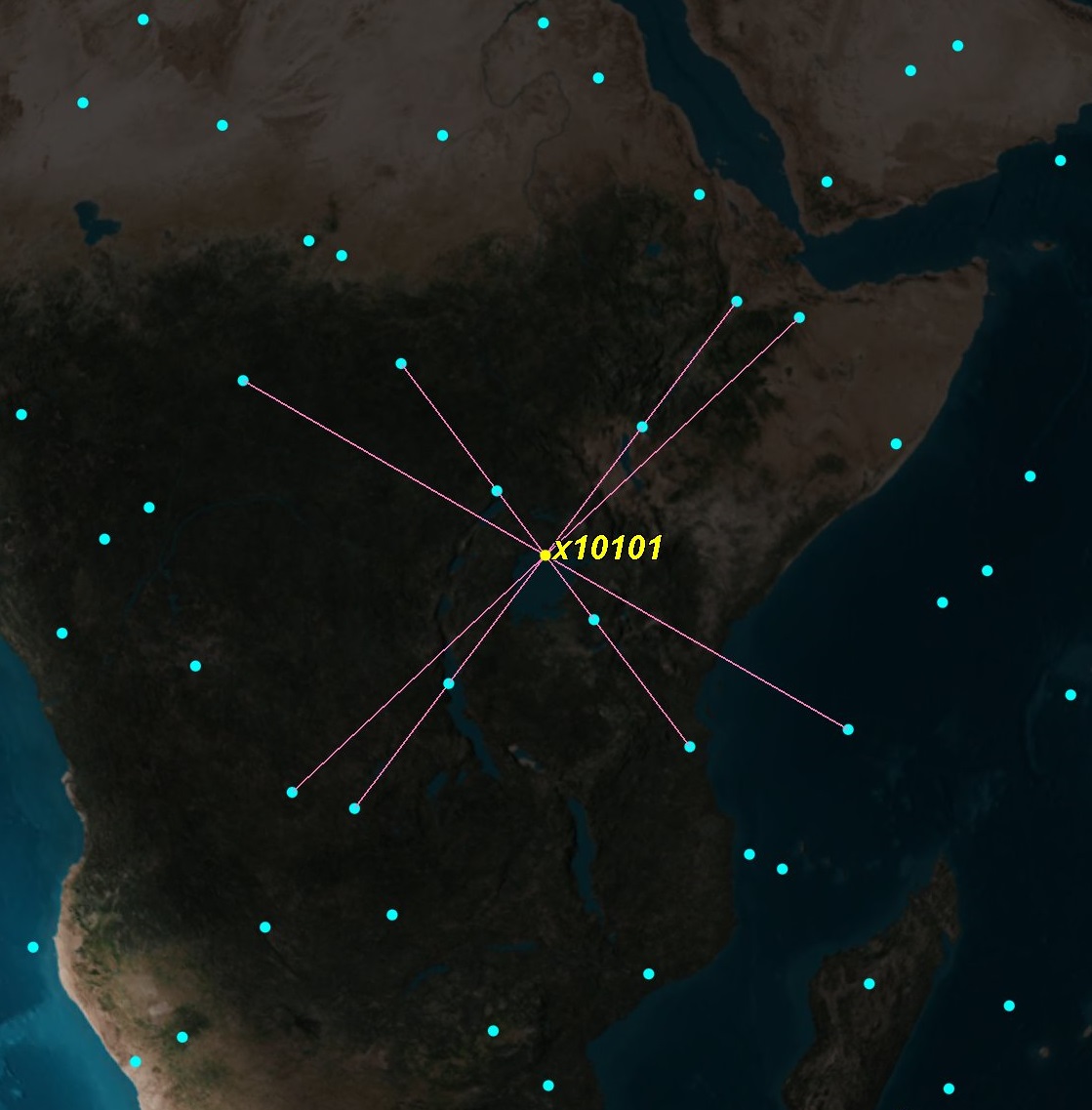}}
	\renewcommand\thefigure{11}\caption{The satellite connectivity for \emph{x10101} at the 1,500 km LISL range with PLs and TLs at $0\degree$ latitude is shown in this figure. The number of possible PLs and TLs that this satellite can establish at this latitude and range is 12, as exhibited in this figure.}
\end{figure}

\begin{figure}[htbp]
	\centerline{\includegraphics[scale=0.297]{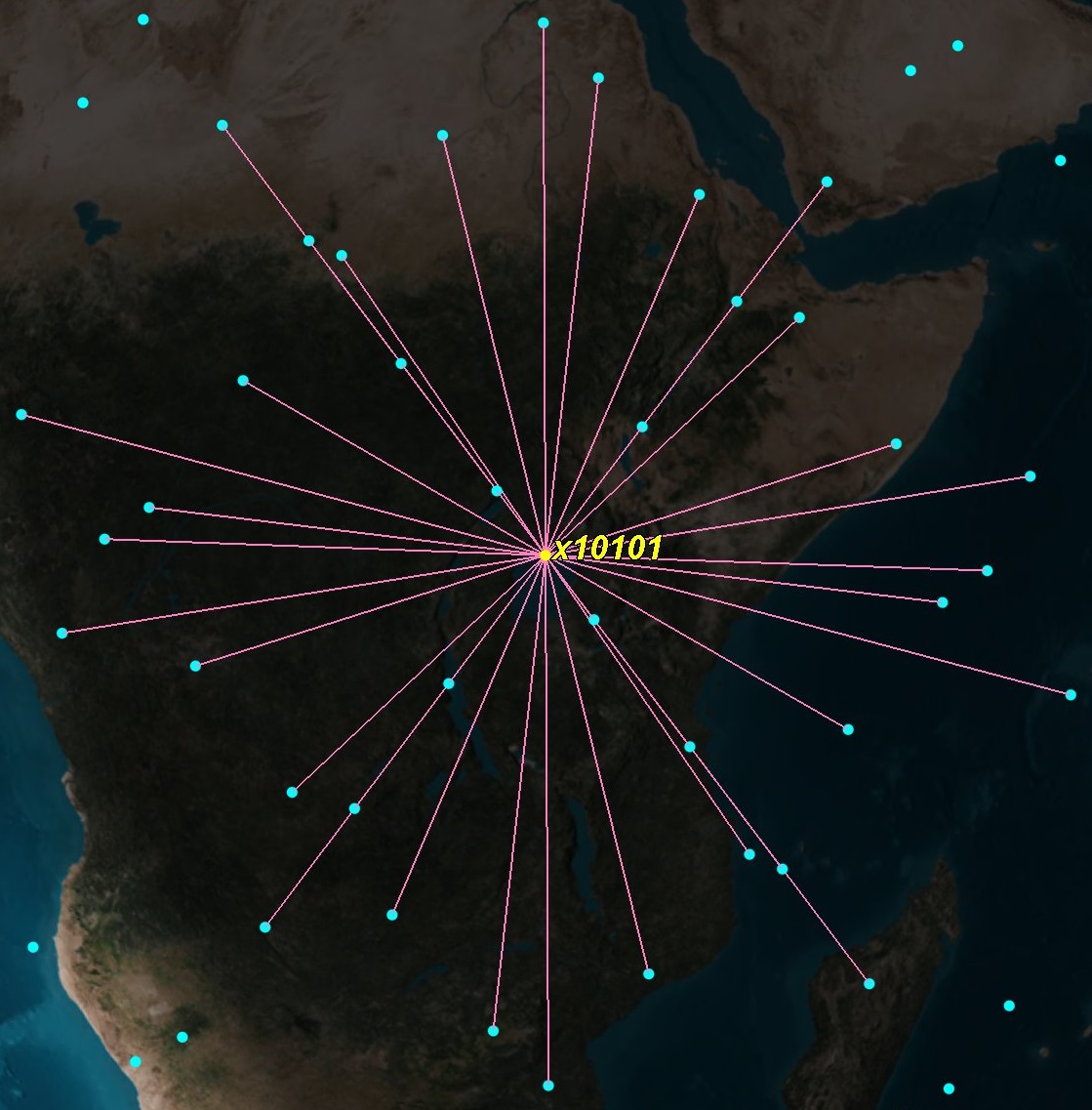}}
	\renewcommand\thefigure{12}\caption{The satellite connectivity for \emph{x10101} at the 2,500 km LISL range with PLs and TLs at $0\degree$ latitude is shown in this figure. The number of possible PLs and TLs that this satellite can establish at this latitude and range is 38, as indicated in this figure.}
\end{figure}

\par At every second or time slot, we take a snapshot of the NG- and NNG-FSOSN for this constellation to study their network connectivity (i.e., the number of possible LISLs as well as the number of possible laser links between satellites and ground stations within the FSOSN at that time slot). Table 3 shows a comparison of the network connectivity at the first time slot for the NG-FSOSN versus the NNG-FSOSN. The network connectivity in the NNG-FSOSN (with PLs and TLs) is at least twice that of the NG-FSOSN (with only PLs). For example, the network connectivity at the 659.5 km LISL range in the NNG-FSOSN is 10,444, and it is around 2.2 times that in the NG-FSOSN, which is 4,756 laser links that are possible within this FSOSN at this range.

\begin{table}
	\centering
	\renewcommand\thetable{3}\caption{Network Connectivity at First Time Slot -- NG-FSOSN vs. NNG-FSOSN.}
	\begin{tabular}{|l|c|c|c|} 
		\hline
		\multirow{2}{*}{\begin{tabular}[c]{@{}l@{}}\textbf{LISL Range}\\\textbf{(km)}\end{tabular}} & \multicolumn{3}{c|}{\textbf{Number of possible laser links in the FSOSN}}  \\ 
		\cline{2-4}
		& \textbf{NG-FSOSN} & \textbf{NNG-FSOSN} & \textbf{Improvement}    \\ 
		\hline
		659.5                                                                                         & 4,756               & 10,444               & 5,688                     \\ 
		\hline
		1,319                                                                                         & 7,932               & 31,176               & 23,244                    \\ 
		\hline
		1,500                                                                                         & 11,100              & 38,784               & 27,684                    \\ 
		\hline
		1,700                                                                                         & 17,436              & 48,180               & 30,744                    \\ 
		\hline
		2,500                                                                                         & 30,108              & 94,116               & 64,008                    \\ 
		\hline
		3,500                                                                                         & 68,124              & 175,788              & 107,664                    \\ 
		\hline
		5,016                                                                                         & 140,998             & 335,928              & 194,930                   \\
		\hline
	\end{tabular}
\end{table} 

\par While studying the effect of TLs on network connectivity at a specific LISL range, we assume the same LISL range for all satellites within an FSOSN. Similar to satellite connectivity (Table 2), network connectivity increases in accordance with an increase in LISL range for both FSOSNs (Table 3). As the LISL range of satellite \emph{x10101} increases, more and farther satellites become available to it for connectivity. Consequently, an increase in LISL range for all satellites in an FSOSN translates into increased network connectivity.

\section{Effect of Temporary Laser Inter-Satellite Links on Network Latency}

\par We use the well-known satellite constellation simulator Systems Tool Kit (STK) Version 12.1 \cite{b17} to simulate the two types of FSOSNs (i.e., an NG-FSOSN, which has only PLs, and an NNG-FSOSN, which has PLs and TLs) based on Starlink’s Phase I constellation to study the effect of TLs on their network latency. While simulating an FSOSN, we generate different IDs for the 1,584 satellites in this constellation. 

\par In the simulation scenario in STK, we construct the satellite constellation for Phase I of Starlink, place ground stations (GSs), specify the LISL range for satellites and range for GSs, and generate all possible links between satellites (PLs for the NG-FSOSN and PLs and TLs for the NNG-FSOSN) and between satellites and GSs. Subsequently, we extract the data of an FSOSN from STK into Python that includes the positions of satellites and GSs, links between satellites, links between satellites and GSs, and duration of the existence of these links. This data is discretized, and it then contains all links that exist at a time slot as well as the positions of satellites at that time slot. This discretized data is used to calculate the length and propagation delay of all links that exist at a time slot. Finally, the NetworkX library in Python \cite{b17a} is used to find the shortest path between GSs for an inter-continental connection over an FSOSN at each time slot.

\par To calculate the shortest paths (in terms of latency) between GSs in cities for different inter-continental connections and different LISL ranges, we use Dijkstra’s algorithm \cite{b18} within the NetworkX library. A time slot gives a snapshot of an FSOSN at that time. We use links between satellites and links between satellites and GSs at a time slot to construct a connectivity graph for an FSOSN at that time slot. Then, we use link propagation delays and node delays in an FSOSN as the weights of the edges and vertices of the connectivity graph, respectively. Next, we employ Dijkstra’s algorithm to find the shortest path over an FSOSN in terms of latency. To this end, we use the connectivity graph as input for this algorithm to calculate the shortest path over an FSOSN between the GSs in cities that minimizes the total weight of the edges and vertices on the path (i.e., the total latency of the path) between the source and destination GSs.

\par Dijkstra's algorithm is used to calculate the shortest path between source and destination ground stations for an inter-continental connection over the two FSOSNs at each time slot. Although the shortest path problem between a single source and single destination can be formulated as a mathematical program, this is not needed since Dijkstra’s algorithm can provide an optimal solution for the single-source shortest path problem \cite{b18a1}.

\par The worst-case computational complexity of Dijkstra’s algorithm to calculate a shortest path over a network is $O(N^2)$ \cite{b18a1} or $O(E\text{ log }N)$ \cite{b18a2}, where $N$ is the number of nodes in the network, and $E$ is the number of edges in the network. However, $O(E\text{ log }N)$ is a tighter bound on the worst-case computational complexity of Dijkstra’s algorithm. For an FSOSN resulting from Starlink’s Phase I constellation consisting of 1,584 satellites and two ground stations (i.e., one source and one destination ground station), the worst-case computational complexity for calculating a shortest path using Dijkstra’s algorithm in terms of nodes is $O(1,586^2)$. 

\par Table 3 shows the network connectivity (i.e., the number of possible links between the nodes in an FSOSN) of the two FSOSNs at different LISL ranges. Due to the additional connectivity arising from TLs, the network connectivity of the NNG-FSOSN is at least twice that of the NG-FSOSN. In terms of nodes as well as edges, for example, the tighter worst-case computational complexity of Dijkstra’s algorithm is $O(335,928\text{ log }1,586)$ and $O(140,998\text{ log }1,586)$ for the NNG-FSOSN and NG-FSOSN, respectively, at the 5,016 km LISL range, and in general, the worst-case computational complexity of calculating a shortest path over the NNG-FSOSN is at least twice that of the NG-FSOSN.

\par The network latency of an FSOSN is the latency of the shortest path, which is equal to the propagation delays of the links (including GS-to-satellite laser link, satellite-to-satellite laser links, and satellite-to-GS laser link) and the node delays of the satellites on the shortest path. To calculate the propagation delay of a laser link, we divide the length of that link by the speed of light in vacuum, i.e., 299,792,458 m/s \cite{ref18b}. The processing delay, which occurs due to on-board routing and switching of packets at a satellite, is considered as 1 ms in \cite{b19}. The queueing delay is minimal for congestion-free networks \cite{b19a}, and the transmission delay is negligible for high-data rate links. Due to assumptions of a congestion-free FSOSN having LISLs with very high data rates, the queueing and transmission delays are considered to be small and are assumed as included in the per hop node delay, and we assume a value of 10 ms for the node delay per satellite/hop to encompass these three delays.

\par FSOSNs can enable low-latency long-distance inter-continental data communications between financial stock markets around the globe, and for different inter-continental connection scenarios, we assume the ground stations in different cities to be located at the financial stock markets within these cities. The range of the ground stations is set to 1,000 km.

\par The shortest path over an FSOSN and/or its latency change at every second due to the high orbital speed of satellites \cite{b9a}, and thereby the duration of a time slot is set as one second to capture these changes. The simulation is run for one hour or 3,600 time slots and a shortest path between the two GSs is calculated over an FSOSN at each time slot for a scenario at a certain LISL range. For example, Figs. 13 and 14 show the shortest paths between GSs over the NG-FSOSN and NNG-FSOSN, respectively, for the Sydney--Sao Paulo inter-continental connection at the 2,500 km LISL range at the same time slot. In these figures, the shortest paths are shown in yellow, and the satellites on the shortest paths are marked in pink. Note that the shortest path for the NNG-FSOSN (with PLs and TLs) consists of seven satellites (or hops) while the one for the NG-FSOSN (with only PLs) comprises nine hops. 

\begin{figure*}[htbp]
	\centerline{\includegraphics[scale=0.405]{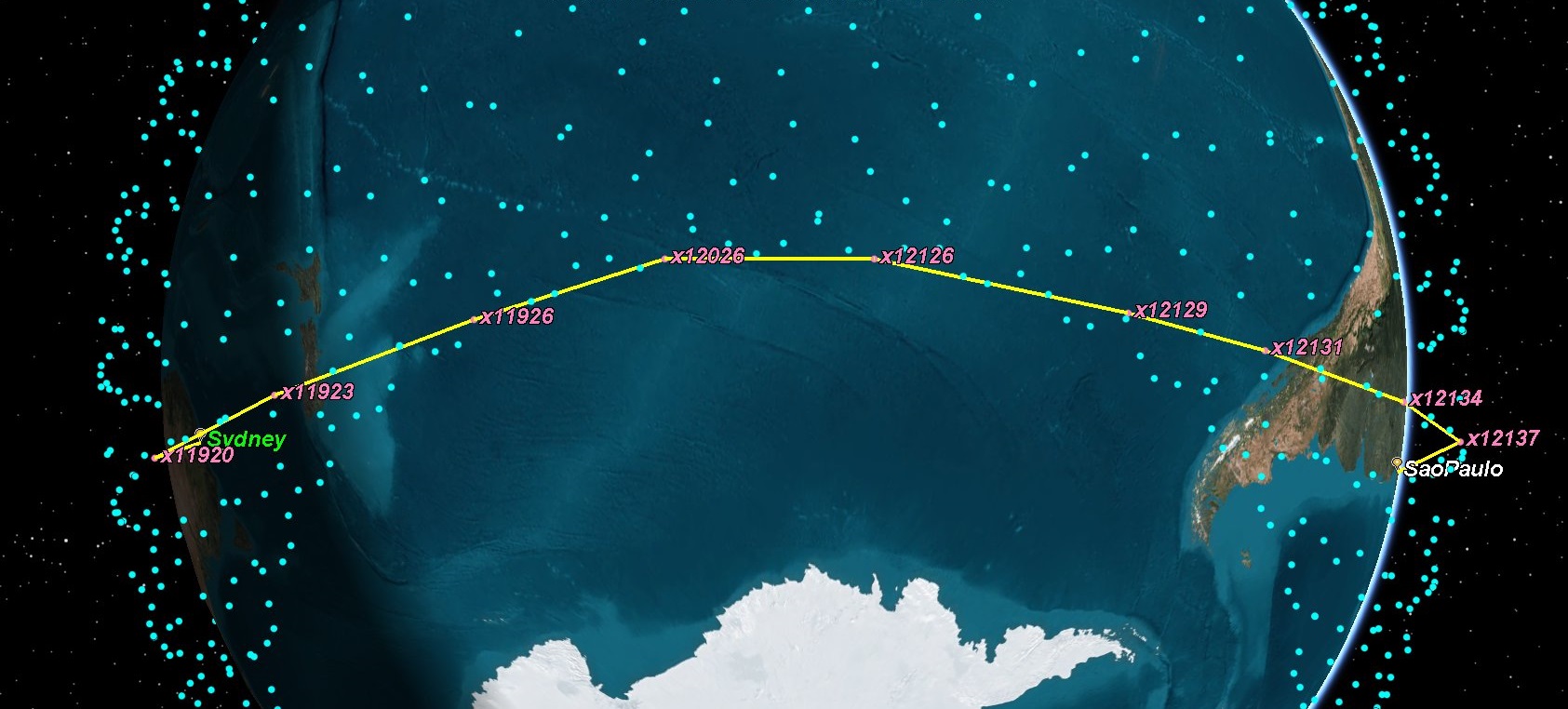}}
	\renewcommand\thefigure{13}\caption{A shortest path for Sydney--Sao Paulo inter-continental connection at the 2,500 km LISL range over the NG-FSOSN is shown in this figure. The shortest path is shown in yellow and the satellites on the shortest path and their markers are displayed in pink. This shortest path over the NG-FSOSN (with only PLs) consists of nine satellites (or hops). The propagation delay, node delay, and latency of this path are 52.24 ms, 90 ms, and 142.24 ms, respectively. The propagation delay of this path is the sum of the propagation delays of all links on this path (i.e., uplink (which is the link between ground station at Sydney and satellite \emph{x11920}), eight LISLs, and downlink [which is the link between satellite \emph{x12137} and ground station at Sao Paulo]); the node delay of this path is the sum of the node delays of all nine hops on this path; and the latency of this path is the sum of its propagation delay and its node delay. Out of all available paths over the NG-FSOSN between Sydney and Sao Paulo at this time slot and this LISL range, this is the shortest path, which means that it provides the lowest latency data communications between Sydney and Sao Paulo over the NG-FSOSN at this time slot and this range.}
\end{figure*}

\begin{figure*}[htbp]
	\centerline{\includegraphics[scale=0.405]{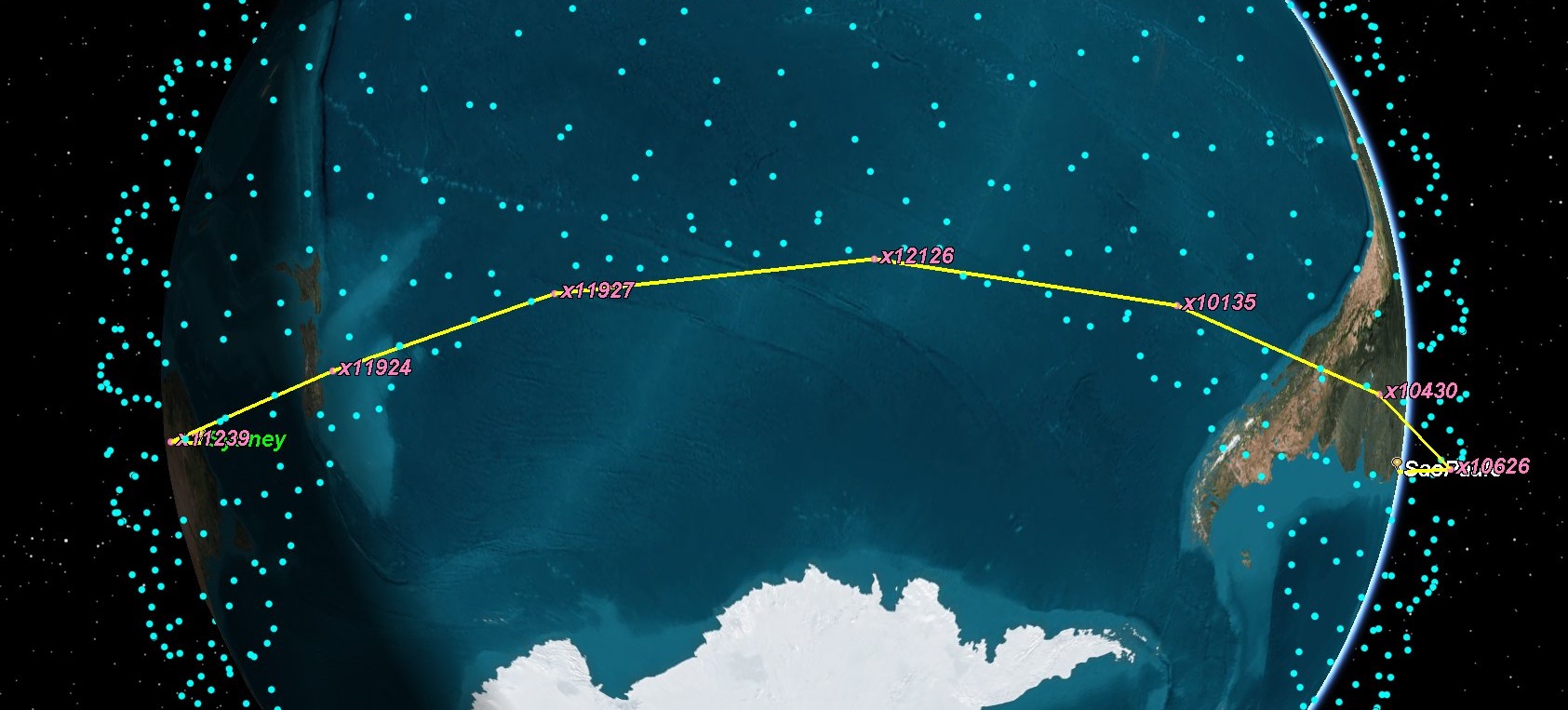}}
	\renewcommand\thefigure{14}\caption{A shortest path for Sydney--Sao Paulo inter-continental connection at the 2,500 km LISL range over the NNG-FSOSN is shown in this figure. This shortest path over the NNG-FSOSN (with PLs and TLs) consists of seven satellites (or hops). The propagation delay, node delay, and latency of this path are 51.21 ms, 70 ms, and 121.21 ms, respectively. Using PLs and TLs in the NNG-FSOSN causes longer and more direct LISLs and fewer hops on this path than that using only PLs in the NG-FSOSN, and this leads to a shortest path with a lower latency. The shortest paths shown in Figs. 13 and 14 for this inter-continental connection at this range over the NG-FSOSN and NNG-FSOSN, respectively, are calculated at the same time slot.}
\end{figure*}

\subsection{Effect of Temporary Laser Inter-Satellite Links on Network Latency for Different LISL Ranges}

\par Table 4 shows a comparison of the average network latency and average number of hops for the NG-FSOSN versus NNG-FSOSN at different LISL ranges for the Sydney--Sao Paulo inter-continental connection. The value of the average network latency and average number of hops at a specific LISL range for an FSOSN in this table is the average of the latencies of the shortest paths at all time slots and the average of the number of hops on the shortest paths at all time slots, respectively.

\par At the 659.5 km LISL range, the average network latency and average number of hops for the two FSOSNs could not be compared. Due to very low satellite connectivity at this LISL range in the NG-FSOSN (i.e., the FSOSN with only PLs and without TLs) that is limited to only two intra-OP neighbors, no shortest path was available between Sydney and Sao Paulo for all 3,600 time slots for this FSOSN, as indicated in Table 4. On the other hand, no shortest path could be found for 1,497 time slots for the NNG-FSOSN (i.e., the FSOSN with PLs and TLs) at this LISL range, and the values shown in Table 4 are calculated for only those 2,103 time slots where shortest paths were found. Sufficient network connectivity is not available at this range for both FSOSNs, and shortest paths could not be ensured at all time slots. Hence, this range is not a feasible LISL range for either of the two FSOSNs for this constellation.

\par At the LISL range of 1,319 km, the satellite connectivity in the NG-FSOSN is restricted to only four intra-OP neighbors, which is also not enough, and no shortest path could be found at any of the time slots for this FSOSN, as shown in Table 4. However, shortest paths were obtained at all times slots for the NNG-FSOSN at this range, and the corresponding values for the average latency (or average network latency) and average number of hops are provided in Table 4. TLs are crucial in ensuring enough network connectivity to guarantee shortest paths at all time slots at this LISL range.

\par For LISL ranges of 1,500 km or higher, there was enough network connectivity in the NG-FSOSN to ensure a shortest path at all time slots. However, the average network latency of the NNG-FSOSN is found to be consistently lower than that of the NG-FSOSN, as we can see in Table 4. The average number of hops for the NNG-FSOSN is found to be either less or equal to that of the NG-FSOSN. For example, at the 5,016 km LISL range, the average number of hops (and thereby node delays) is the same for both FSOSNs; however, better shortest paths consisting of more direct LISLs result in lower propagation delays and lower average latency for the NNG-FSOSN. The average network latency and average number of hops of the two FSOSNs for the Sydney--Sao Paulo inter-continental connection are also plotted in Fig. 15 for the 1,500 km--5,016 km LISL ranges so that trends are apparent.

\begin{table*}
	\centering
	\renewcommand\thetable{4}\caption{Average Network Latency, Average Number of Hops, and Number of Time Slots Where a Shortest Path is Available for the Sydney--Sao Paulo Inter-Continental Connection at Different LISL Ranges -- NG-FSOSN vs. NNG-FSOSN.}
	\begin{tabular}{|l|c|c|c|c|c|c|c|c|} 
		\hline
		\multirow{2}{*}{\begin{tabular}[c]{@{}l@{}}\textbf{LISL Range} \\\textbf{(km)}\end{tabular}} & \multicolumn{3}{c|}{\textbf{Average Network Latency}}                                                                                                                                                                                 & \multicolumn{3}{c|}{\textbf{Average Number of Hops}}          & \multicolumn{2}{c|}{\begin{tabular}[c]{@{}c@{}}\textbf{Number of Time Slots where}\\\textbf{a shortest path is available}\end{tabular}}  \\ 
		\cline{2-9}
		& \begin{tabular}[c]{@{}c@{}}\textbf{NG-FSOSN} \\\textbf{(ms)}\end{tabular} & \begin{tabular}[c]{@{}c@{}}\textbf{NNG-FSOSN} \\\textbf{(ms)}\end{tabular} & \begin{tabular}[c]{@{}c@{}}\textbf{Improvement} \\\textbf{(ms)}\end{tabular} & \textbf{NG-FSOSN} & \textbf{NNG-FSOSN} & \textbf{Improvement} & \textbf{NG-FSOSN} & \textbf{NNG-FSOSN}                                                                                                    \\ 
		\hline
		659.5                                                                                        & --                                                                         & 299.04                                                                     & --                                                                            & --                 & 24.46              & --                    & 0                 & 2,103                                                                                                                 \\ 
		\hline
		1,319                                                                                        & --                                                                         & 172.33                                                                     & --                                                                            & --                 & 12.00              & --                    & 0                 & 3,600                                                                                                                 \\ 
		\hline
		1,500                                                                                        & 188.44                                                                    & 171.61                                                                     & 16.83                                                                        & 13.54             & 11.94              & 1.60                 & 3,600             & 3,600                                                                                                                 \\ 
		\hline
		1,700                                                                                        & 180.78                                                                    & 157.35                                                                     & 23.43                                                                        & 12.83             & 10.51              & 2.32                 & 3,600             & 3,600                                                                                                                 \\ 
		\hline
		2,500                                                                                        & 142.37                                                                    & 124.17                                                                     & 18.20                                                                        & 9.00              & 7.19               & 1.81                 & 3,600             & 3,600                                                                                                                 \\ 
		\hline
		3,500                                                                                        & 112.11                                                                    & 109.19                                                                     & 2.92                                                                         & 6.00              & 5.75               & 0.25                 & 3,600             & 3,600                                                                                                                 \\ 
		\hline
		5,016                                                                                        & 91.65                                                                     & 90.89                                                                      & 0.76                                                                         & 4.00              & 4.00               & 0.00                 & 3,600             & 3,600                                                                                                                 \\
		\hline
	\end{tabular}
\end{table*}

\begin{figure*}[t!]
	\centering
	\begin{subfigure}[t]{0.5\textwidth}
		\centering
		\includegraphics[width=\textwidth]{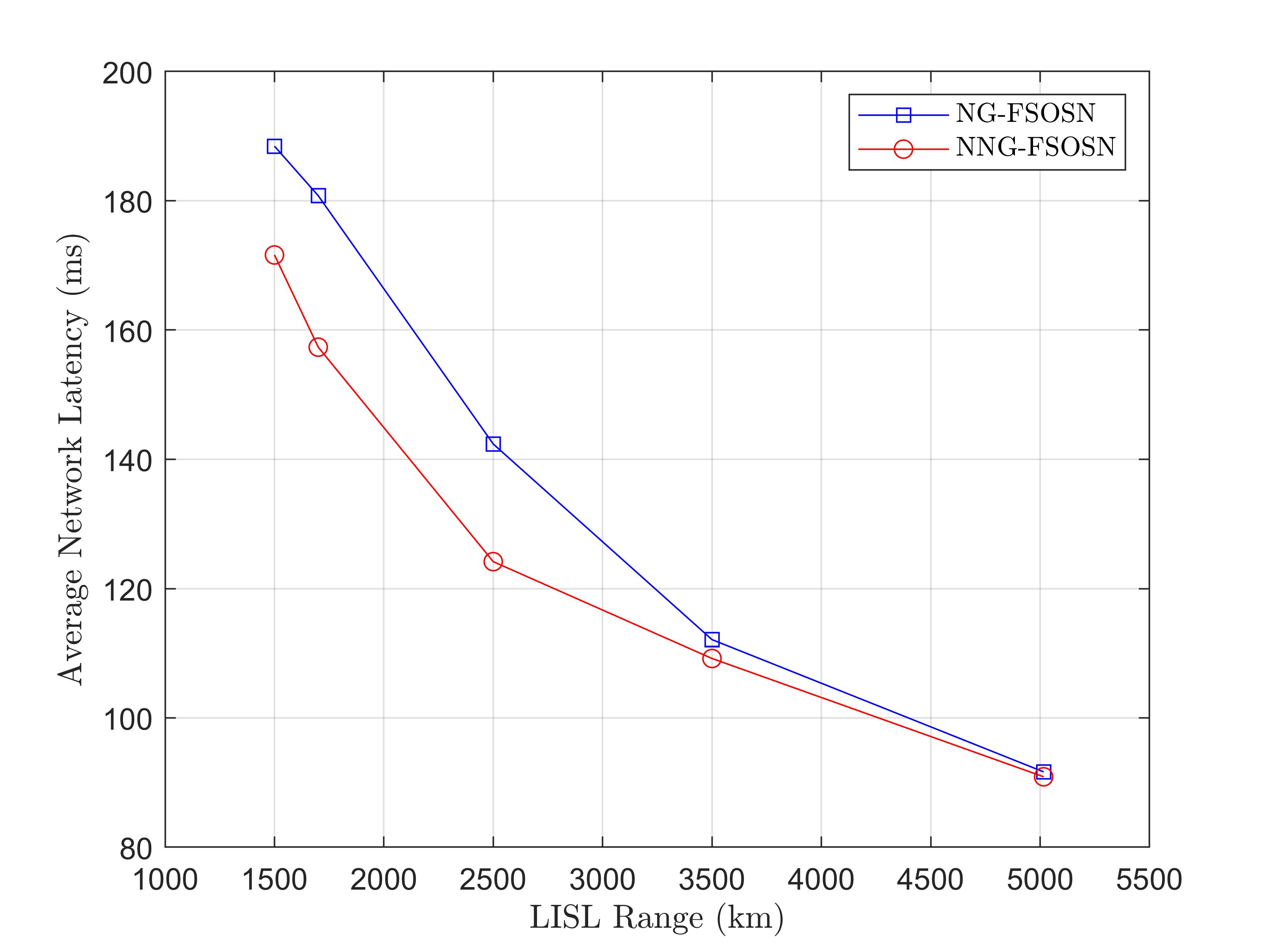}
		\caption{Average Network Latency.}
		%\vspace{0.1cm}
	\end{subfigure}%
	%\hfill
	%\par\bigskip
	\begin{subfigure}[t]{0.5\textwidth}
		\centering
		\includegraphics[width=\textwidth]{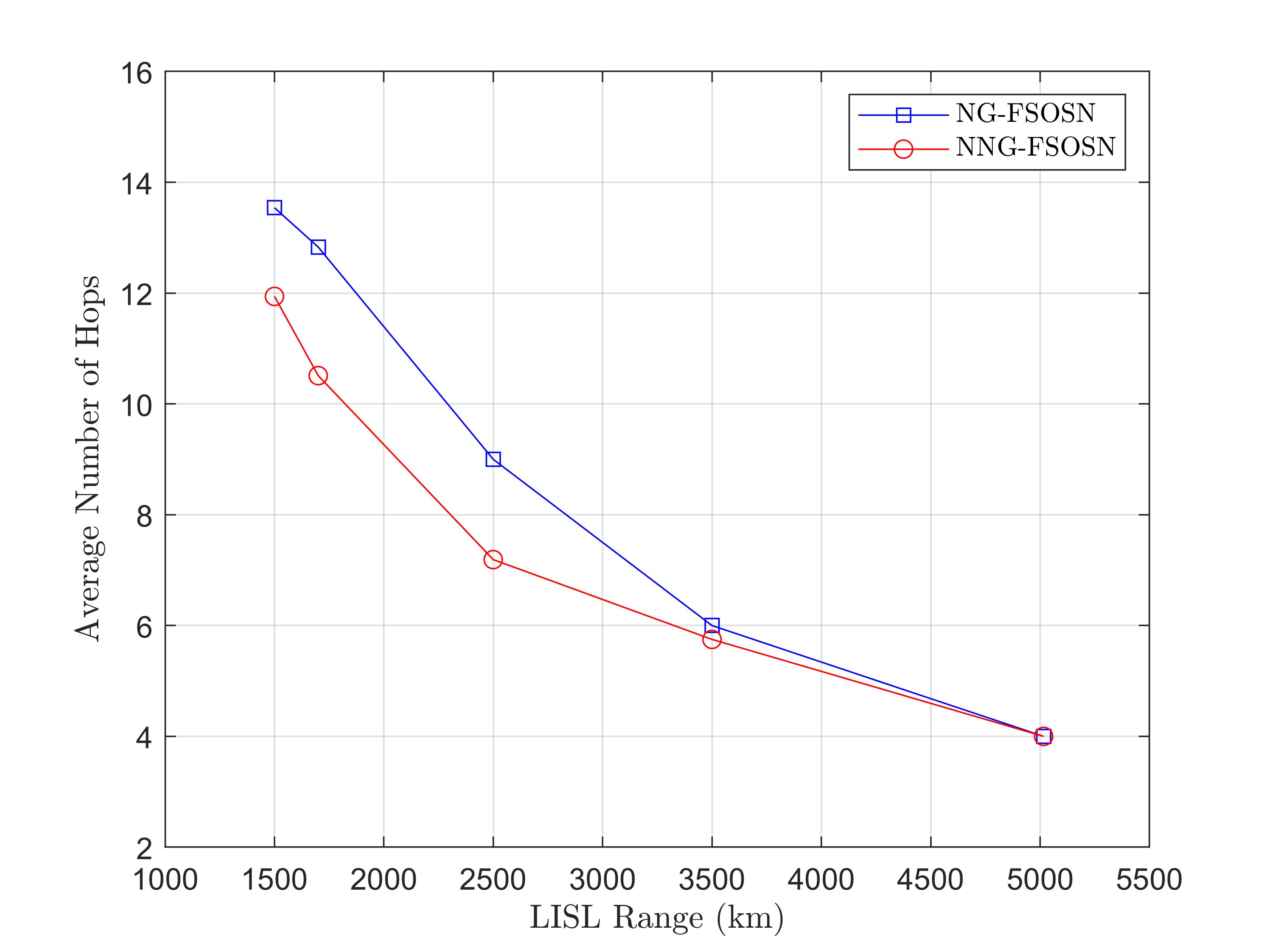}
		\caption{Average Number of Hops.}
	\end{subfigure}
	%\hfill
	\vspace*{0.0001mm}
	\renewcommand\thefigure{15}\caption{The average network latency and average number of hops of the NG-FSOSN vs. NNG-FSOSN for the Sydney-Sao Paulo inter-continental connection are shown in these plots for the 1,500 km--5,016 km LISL ranges. The solid blue lines with square markers and the solid red lines with circle markers represent the NG-FSOSN and NNG-FSOSN, respectively. The NNG-FSOSN (i.e., the FSOSN with PLs and TLs) offers lower average network latency than the NG-FSOSN (i.e., the FSOSN with only PLs) at all LISL ranges. Note that the average number of hops for both networks is the same at the 5,016 km LISL range. However, the NNG-FSOSN still outperforms the NG-FSOSN by 0.76 ms in terms of average network latency.}
\end{figure*}

\subsection{Effect of Temporary Laser Inter-Satellite Links on Network Latency for Different Inter-Continental Connections}

\par We also studied the effect of TLs on the network latency of the two FSOSNs under different inter-continental connection scenarios for two different LISL ranges, specifically 1,700 km and 5,016 km. Figs. 16 and 17 show the results for the average network latency and average number of hops for the two FSOSNs for the Toronto--Istanbul, Madrid--Tokyo, and New York--Jakarta inter-continental connection scenarios at LISL ranges of 1,700 km and 5,016 km, respectively.  

\par At the 1,700 km LISL range, the NNG-FSOSN provides a lesser average number of hops (which translates into lower node delays) and lower average network latency compared to the NG-FSOSN. In all scenarios, the two FSOSNs are found to have the same average number of hops at the 5,016 km LISL range; however, the NNG-FSOSN still performs better than the NG-FSOSN at this range in terms of average network latency.

\begin{figure*}[t!]
	\centering
	\begin{subfigure}[t]{0.5\textwidth}
		\centering
		\includegraphics[width=\textwidth]{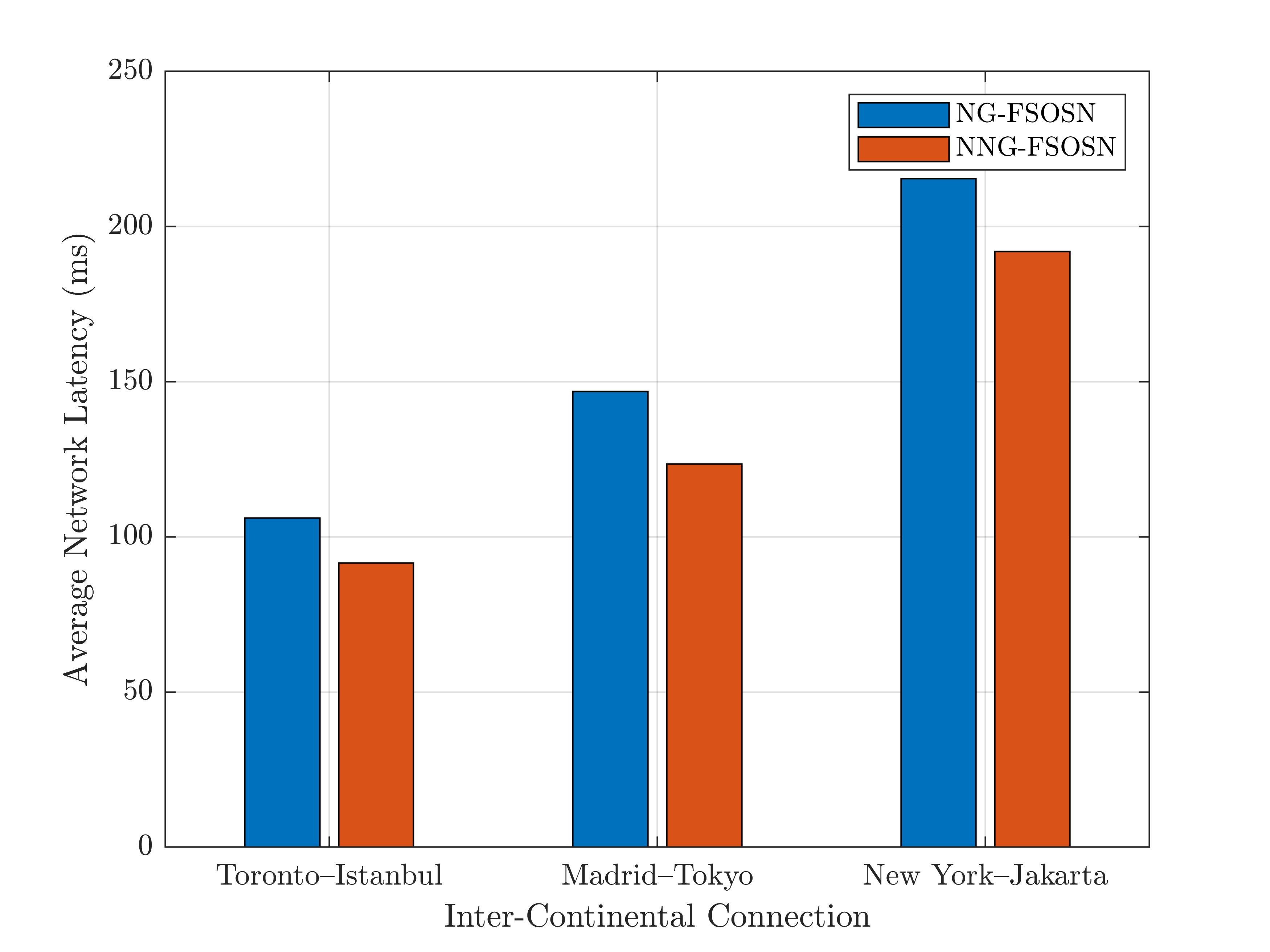}
		\caption{Average Network Latency.}
		%\vspace{0.1cm}
	\end{subfigure}%
	%\hfill
	%\par\bigskip
	\begin{subfigure}[t]{0.5\textwidth}
		\centering
		\includegraphics[width=\textwidth]{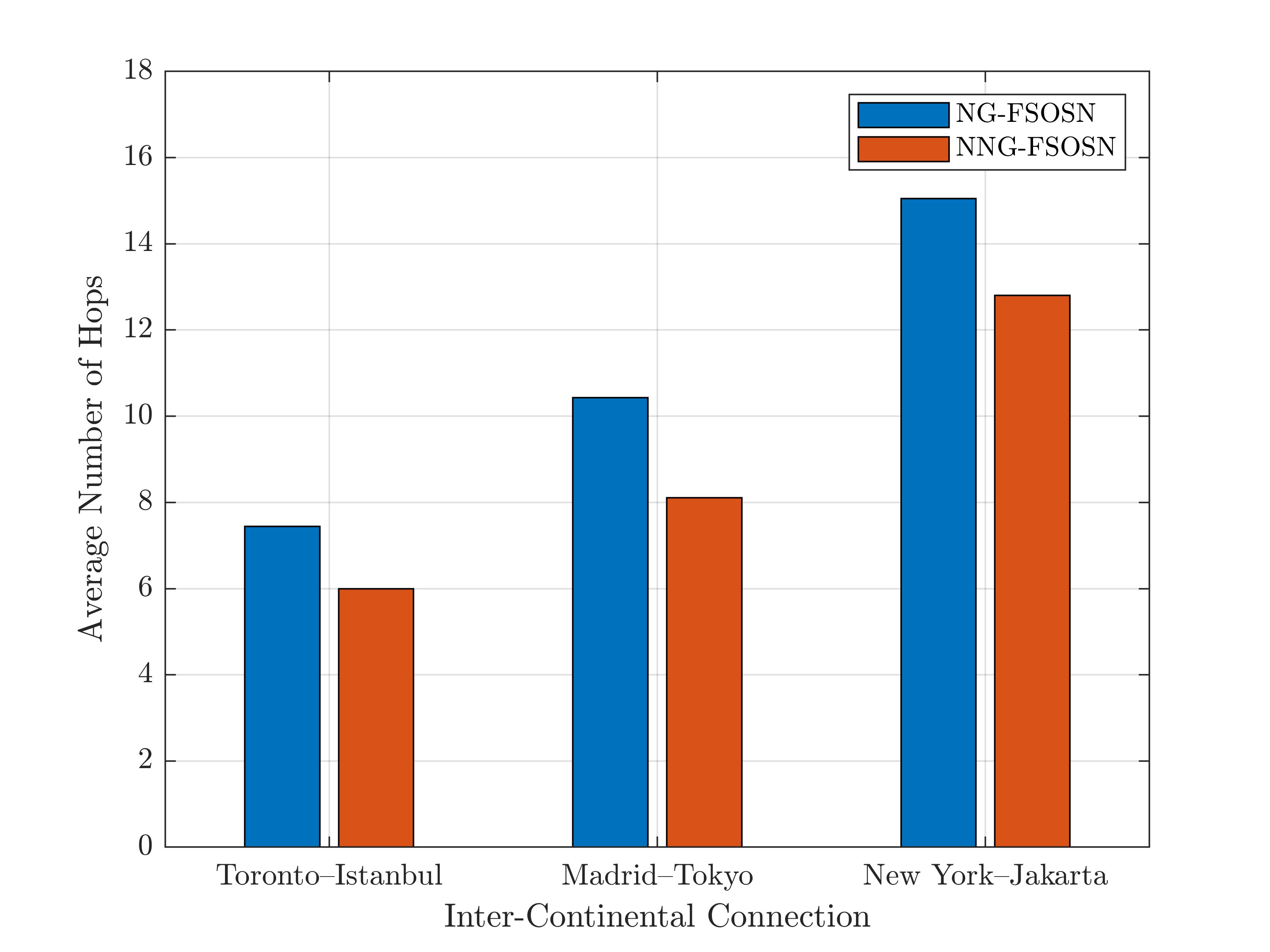}
		\caption{Average Number of Hops.}
	\end{subfigure}
	%\hfill
	\vspace*{0.0001mm}
	\renewcommand\thefigure{16}\caption{The average network latency and average number of hops of the NG-FSOSN vs. NNG-FSOSN for different inter-continental connections at the 1,700 km LISL range are shown in these plots. The blue bars and red bars represent the NG-FSOSN and NNG-FSOSN, respectively. The NNG-FSOSN (i.e., the FSOSN with PLs and TLs) provides lower average network latency than the NG-FSOSN (i.e., the FSOSN with only PLs and without TLs) at this LISL range for all inter-continental connections.}
\end{figure*}

\begin{figure*}[t!]
	\centering
	\begin{subfigure}[t]{0.5\textwidth}
		\centering
		\includegraphics[width=\textwidth]{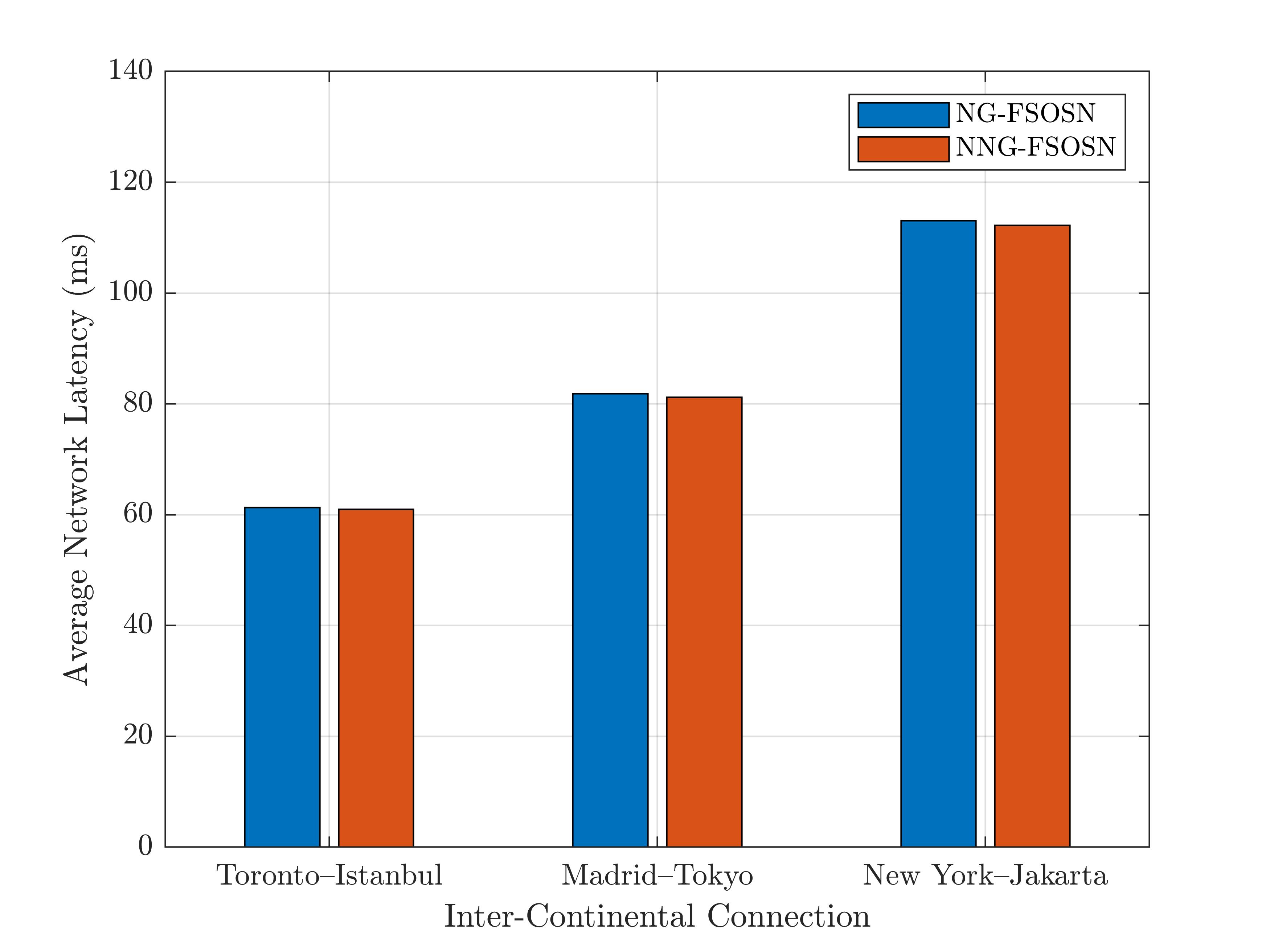}
		\caption{Average Network Latency.}
		%\vspace{0.1cm}
	\end{subfigure}%
	%\hfill
	%\par\bigskip
	\begin{subfigure}[t]{0.5\textwidth}
		\centering
		\includegraphics[width=\textwidth]{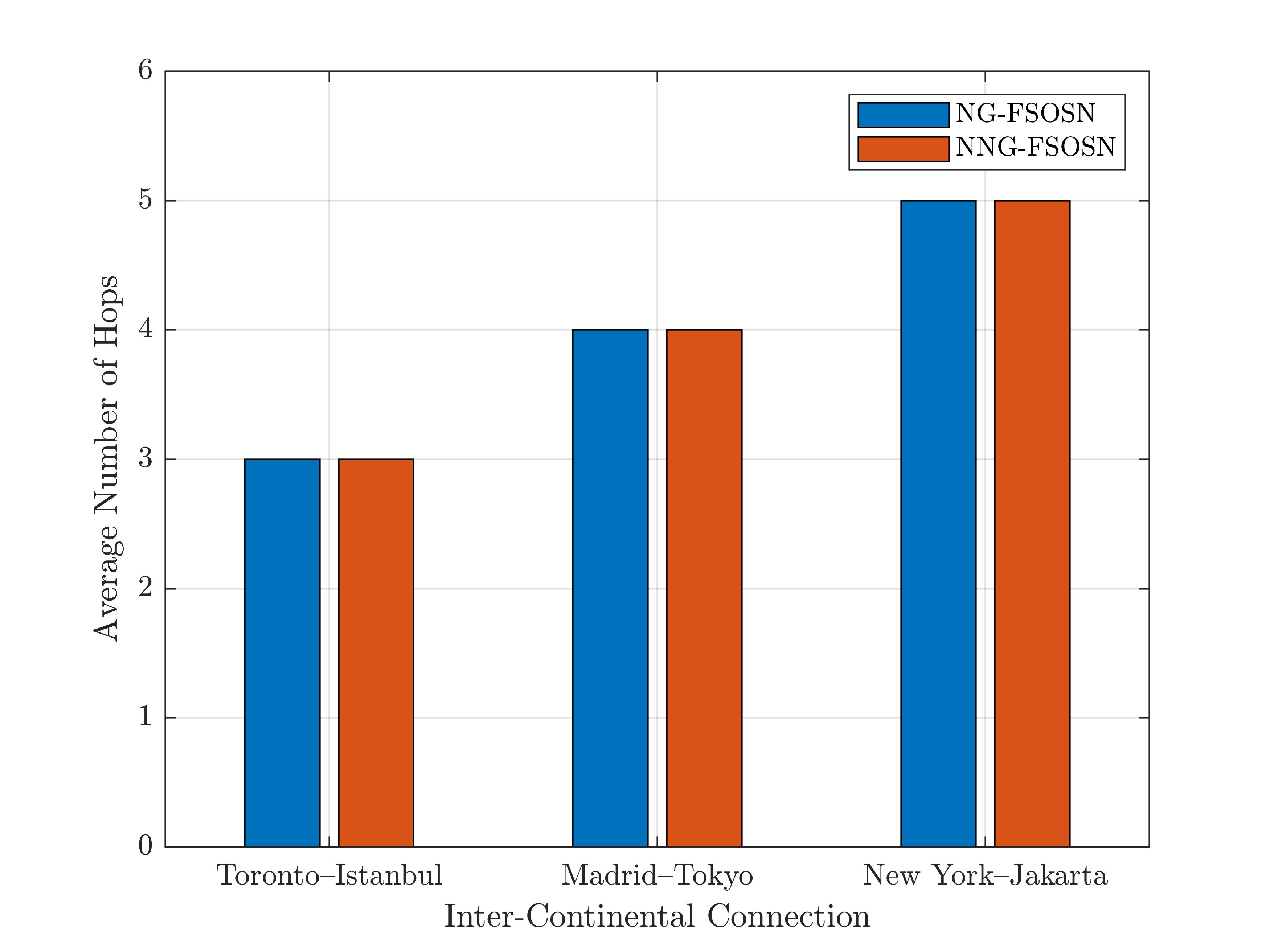}
		\caption{Average Number of Hops.}
	\end{subfigure}
	%\hfill
	\vspace*{0.0001mm}
	\renewcommand\thefigure{17}\caption{The average network latency and average number of hops of the NG-FSOSN vs. NNG-FSOSN for different inter-continental connections at the 5,016 km LISL range are shown in these plots. The blue bars and red bars represent the NG-FSOSN and NNG-FSOSN, respectively. The NNG-FSOSN (i.e., the FSOSN with PLs and TLs) delivers lower average network latency than the NG-FSOSN (i.e., the FSOSN with only PLs) at this LISL range for all inter-continental connections. Although the two FSOSNs have the same average number of hops at this LISL range for all inter-continental connections, the NNG-FSOSN still outperforms the NG-FSOSN in terms of average network latency by 0.33 ms, 0.64 ms, and 0.88 ms for the Toronto--Istanbul, Madrid--Tokyo, and New York--Jakarta inter-continental connections, respectively.}
\end{figure*}

\subsection{Discussion}

\par At LISL ranges of 1,500 km, 1,700 km, and 2,500 km in the Sydney--Sao Paulo inter-continental connection scenario (Fig. 15) as well as at the 1,700 km LISL range in the Toronto--Istanbul, Madrid--Tokyo, and New York--Jakarta inter-continental connection scenarios (Fig. 16), the NNG-FSOSN performs significantly better than the NG-FSOSN in terms of average network latency by providing a lower average number of hops, resulting in lower node delays. At the LISL range of 3,500 km for Sydney--Sao Paulo (Fig. 15), the average number of hops is nearly same for the NNG-FSOSN and NG-FSOSN, and the improvement in average network latency with the NNG-FSOSN is small. At the 5,016 km LISL range in all four inter-continental connection scenarios (Figs. 15 and 17), the average number of hops is same for the NNG-FSOSN and NG-FSOSN. Notably, the NNG-FSOSN still performs better than the NG-FSOSN in terms of average network latency. This is due to better shortest paths (consisting of PLs as well as TLs) with lower propagation delays resulting in lower average latency for the NNG-FSOSN. The results in Figs. 15--17 clearly indicate the importance of TLs in achieving lower network latency.

\par The NNG-FSOSN, which employs PLs and TLs, provides an improvement in the average network latency of more than 16 ms compared to the NG-FSOSN, which has no TLs, at LISL ranges of 1,500 km, 1,700 km, and 2,500 km for the Sydney--Sao Paulo inter-continental connection. However, this improvement in average network latency decreases to less than 3 ms and 1 ms at the 3,500 km and 5,016 km LISL ranges, respectively, for this inter-continental connection. We observe an improvement of approximately 15 ms for the Toronto--Istanbul inter-continental connection and more than 23 ms for the Madrid--Tokyo and New York--Jakarta inter-continental connections at the 1,700 km LISL range. This improvement falls below 1 ms for these inter-continental connections at the 5,016 km LISL range. This indicates that the LISL range affects the improvement in average network latency that is achieved by using TLs, and this improvement is significant at LISL ranges of 1,500 km, 1,700 km, and 2,500 km.

\par At LISL ranges of 1,700 km and 5,016 km, the results in Figs. 16 and 17 show that the average network latency for both FSOSNs increases in accordance with the length of the inter-continental distance between GSs in cities within a scenario. Note that the terrestrial distance between GSs for the Toronto--Istanbul, Madrid--Tokyo, and New York--Jakarta inter-continental connections along the surface of the Earth can be calculated as 8,198 km, 10,778 km, and 16,198 km, respectively, based on the coordinates (i.e., latitudes and longitudes) of the stock exchanges in these cities, assuming the radius of the Earth is 6,378 km. The longer the terrestrial distance between GSs within a scenario, the higher the latencies of the shortest paths over an FSOSN, and thereby the higher the average network latency for that scenario. 

\par As with Fig. 15, we can see in Figs. 16 and 17 that the higher the LISL range for a scenario, the lower the average network latency of an FSOSN. For example, for the Toronto--Istanbul inter-continental connection scenario, the average latency for the NNG-FSOSN at the 1,700 km and 5,016 km LISL ranges is 91.52 ms and 60.95 ms, respectively. As the LISL range increases, the satellite connectivity and network connectivity increase (Tables 2 and 3), which results in better shortest paths consisting of longer LISLs with farther neighbors leading to fewer LISLs (lower propagation delays), fewer hops (lower node delays), and lower average network latency.  

%\par \vspace{0.5cm}

\subsection{Design Guidelines}
In the following, we list some design guidelines for the two FSOSNs that are realized from Starlink's Phase I constellation:
\begin{itemize}
	\item Due to insufficient network connectivity with or without TLs in the NNG-FSOSN or NG-FSOSN, respectively, the LISL range of 659.5 km is not a feasible LISL range for an FSOSN based on Starlink’s Phase I constellation. 
	\item The 1,319 km LISL range is also an unfeasible range without TLs, and TLs are required to ensure the shortest paths at all time slots at this range. 
	\item For a 1,500 km LISL range or higher, shortest paths are available at all time slots in the NNG-FSOSN (with TLs) as well as in the NG-FSOSN (without TLs).
	\item A significant part of the network latency of a shortest path over an FSOSN depends on the number of hops on the shortest path and thereby the node delay of the shortest path (i.e., sum of the node delays of all hops or satellites on the shortest path).
	\item The network latency of a shortest path over an FSOSN for an inter-continental connection scenario is proportional to the terrestrial distance between ground stations in that scenario.
	\item A higher LISL range for satellites in an FSOSN can provide a lower average network latency.
	\item The improvement in average network latency with TLs is affected by the LISL range. 
	\item LISL ranges of 1,500 km, 1,700 km, and 2,500 km are more suitable to leverage TLs in an FSOSN resulting from Starlink’s Phase I constellation as these LISL ranges provide a significant improvement in average network latency for the NNG-FSOSN compared to the NG-FSOSN.
\end{itemize}

\section{Conclusions}

\par In this work, we investigated the impact of temporary LISLs on network connectivity and network latency of FSOSNs under different scenarios for long-distance inter-continental data communications and different LISL ranges for satellites. To do this, we used the satellite constellation for Phase I of Starlink. Due to higher satellite connectivity and network connectivity in the NNG-FSOSN resulting from TLs, better shortest paths between GSs in cities over this FSOSN were found offering lower propagation and node delays and lower network latency. This resulted in lower average network latency for the NNG-FSOSN (i.e., the FSOSN with PLs and TLs) compared to that for the NG-FSOSN (i.e., the FSOSN with only PLs). The improvement in average network latency was seen to be significant at LISL ranges of 1,500 km, 1,700 km, and 2,500 km, where it was 16.83 ms, 23.43 ms, and 18.20 ms, respectively, for the Sydney--Sao Paulo inter-continental connection. For the Toronto--Istanbul, Madrid--Tokyo, and New York--Jakarta inter-continental connections with an LISL range of 1,700 km, it was 14.58 ms, 23.35 ms, and 23.52 ms, respectively.

\par Due to high LISL setup times of LCTs available for satellites in NG-FSOSNs, TLs are currently considered undesirable in these satellite networks. However, TLs are likely to become feasible among satellites in NNG-FSOSNs, where we envision LCTs to become available through technological advancements that will offer instantaneous setup of LISLs. In addition to being beneficial for intra-FSOSN data communications, TLs will also be needed to communicate between satellites in FSOSNs at different altitudes and will be essential for inter-FSOSN data communications where permanent LISLs are not possible. The delay incurred in establishing an LISL between a pair of satellites can be referred to as LISL setup delay, and in future we plan to study this delay in order to quantify its impact on network latency of NNG-FSOSNs.

\section*{Acknowledgment}

\par The authors would like to thank AGI for the Systems Tool Kit (STK) platform.

\begin{IEEEbiographynophoto}{Aizaz U. Chaudhry} (M’10–SM’20) received the B.Sc. degree in Electrical Engineering from the University of Engineering and Technology Lahore, Lahore, Pakistan, in 1999, and the M.A.Sc. and Ph.D. degrees in Electrical and Computer Engineering from Carleton University, Ottawa, Canada, in 2010 and 2015, respectively. 

He is currently a Senior Research Associate with the Department of Systems and Computer Engineering at Carleton University. Previously, he worked as an NSERC Postdoctoral Research Fellow at Communications Research Centre Canada, Ottawa. His research work has been published in refereed venues, and has received several citations. He has authored and co-authored more than thirty-five publications. His research interests include the application of machine learning and optimization in wireless networks.

Dr. Chaudhry is a licensed Professional Engineer in the Province of Ontario, a Senior Member of IEEE, and a Member of IEEE ComSoc's Technical Committee on Satellite and Space Communications. He serves as a technical reviewer for conferences and journals on a regular basis. He has also served as a TPC Member for conferences, such as IEEE ICC 2021 Workshop 6GSatComNet, IEEE ICC 2022 Workshop 6GSatComNet, IEEE VTC 2022-Spring, IEEE VTC 2022-Fall, twenty-first international conference on networks (ICN 2022), eleventh advanced satellite multimedia systems conference (ASMS 2022), seventeenth signal processing for space communications workshop (SPSC 2022), first international symposium on satellite communication systems and services (SCSS 2022), and fifteenth international conference on communication systems and networks (COMSNETS 2023). 
\end{IEEEbiographynophoto}

%\vspace{-1cm}

\begin{IEEEbiographynophoto}{Halim Yanikomeroglu} (F'17) received his Ph.D. degree in Electrical and Computer Engineering from the University of Toronto, Toronto, Canada, in 1998. 
	
He is currently a Full Professor at Carleton University. He contributed to 4G/5G technologies and non-terrestrial networks. His industrial collaborations resulted in 39 granted patents. He supervised or hosted in his lab a total of 140 postgraduate researchers. He co-authored IEEE papers with faculty members in 80+ universities in 25 countries. 

Dr. Yanikomeroglu is a Fellow of IEEE, Engineering Institute of Canada, and Canadian Academy of Engineering, and an IEEE Distinguished Speaker for ComSoc and VTS. He is currently chairing the WCNC Steering Committee, and he is a Member of PIMRC Steering Committee and ComSoc Emerging Technologies Committee. He served as the General Chair of two VTCs and TP Chair of three WCNCs. He chaired ComSoc's Technical Committee on Personal Communications. He received several awards, including ComSoc Wireless Communications TC Recognition Award (2018), VTS Stuart Meyer Memorial Award (2020), and ComSoc Fred W. Ellersick Prize (2021).
\end{IEEEbiographynophoto}

\begin{thebibliography}{00}

\bibitem{b1} SpaceX FCC update, Nov. 2016, ``SpaceX Non-Geostationary Satellite System, Attachment A, Technical Information to Supplement Schedule S,'' [Online]. Available: \url{https://licensing.fcc.gov/myibfs/download.do?attachment_key=1158350}, Accessed: July 11, 2022.

\bibitem{b1a} SpaceX FCC update, Mar. 2017, ``SpaceX V-Band Non-Geostationary Satellite System, Attachment A, Technical Information to Supplement Schedule S,'' [Online]. Available: \url{https://licensing.fcc.gov/myibfs/download.do?attachment_key=1190019}, Accessed: July 11, 2022.

\bibitem{b1b} SpaceX FCC update, Nov. 2018, ``SpaceX Non-Geostationary Satellite System, Attachment A, Technical Information to Supplement Schedule S,'' [Online]. Available: \url{https://licensing.fcc.gov/myibfs/download.do?attachment_key=1569860}, Accessed: July 11, 2022.

\bibitem{b2} Telesat Canada FCC update, Nov. 2016, ``Petition for Declaratory Ruling,'' [Online]. Available: \url{https://fcc.report/IBFS/SAT-PDR-20161115- 00108/1158133.pdf}, Accessed: July 11, 2022.

\bibitem{b2a} Telesat Canada FCC update, May 2020, ``Application for Modification for Market Access Authorization,'' [Online]. Available: \url{https://fcc.report/IBFS/SAT-MPL-20200526-00053/2378318.pdf}, Accessed: July 11, 2022.

\bibitem{b3} R. Jewett, ``SpaceX Touts 100 Mbps Starlink Test Speeds, Confirms Inter-Satellite Links,'' \emph{Via Satellite}, Sep. 2020, [Online]. Available: \url{https://www.satellitetoday.com/launch/2020/09/03/spacex-touts-100-mbps-starlink-test-speeds-confirms-inter-satellite-links/}, Accessed: July 11, 2022.

\bibitem{b4} S. Erwin, ``Thales Alenia selected to build Telesat’s broadband constellation,'' \emph{Space News}, Feb. 2021, [Online]. Available: \url{https://spacenews.com/thales-alenia-selected-to-build-telesats-broadband-constellation/}, Accessed: July 11, 2022.

\bibitem{b5} A.U. Chaudhry and H. Yanikomeroglu, ``Free Space Optics for Next-Generation Satellite Networks,'' \emph{IEEE Consumer Electronics Magazine}, vol. 10(6), pp. 21--31, Nov. 2021.

\bibitem{b6} R. Jewett, ``Latest Starlink Satellites Equipped with Laser Communications, Musk Confirms,'' \emph{Via Satellite}, Jan. 2021, [Online]. Available: \url{https://www.satellitetoday.com/broadband/2021/01/25/latest-starlink-satellites-equipped-with-laser-communications-musk-confirms/}, Accessed: July 11, 2022.

\bibitem{b7} S. Muncheberg, C. Gal, J. Horwath, H. Kinter, L.M. Navajas, and M. Soutullo, ``Development Status and Breadboard Results of a Laser Communication Terminal for Large LEO Constellations,'' in \emph{Proc. Society of Photo-Optical Instrumentation Engineers}, vol. 11180, article id. 1118034, 2019, pp. 1--13.

\bibitem{b8} M. Motzigemba, H. Zech, and P. Biller, ``Optical Inter Satellite Links for Broadband Networks,'' in \emph{Proc. 9th International Conference on Recent Advances in Space Technologies}, Istanbul, Turkey, 2019, pp. 509–512.

\bibitem{b9} A.U. Chaudhry and H. Yanikomeroglu, ``Laser Intersatellite Links in a Starlink Constellation: A Classification and Analysis,'' \emph{IEEE Vehicular Technology Magazine}, vol. 16(2), pp. 48--56, Jun. 2021.

\bibitem{b9a} A.U. Chaudhry and H. Yanikomeroglu, ``Optical Wireless Satellite Networks versus Optical Fiber Terrestrial Networks: The Latency Perspective – Invited Paper,'' in \emph{Proc. 30th Biennial Symposium on Communications}, Saskatoon, Canada, 2021, pp. 1--6, [Online]. Available: \url{https://arxiv.org/abs/2106.07737}, Accessed July 11, 2022.

\bibitem{b9b} R. Martin, ``Wall Street’s Quest to Process Data at the Speed of Light,'' \emph{Information Week}, Apr. 2007, [Online]. Available: \url{https://rubinow.com/2007/04/21/wall-streets-quest-to-process-data-at-the-speed-of-light/}, Accessed: July 11, 2022.

\bibitem{b9c} H. Kaushal, V.K. Jain, and S. Kar, ``Acquisition, Tracking, and Pointing,'' in \emph{Free Space Optical Communication}, New Delhi: Springer-Verlag, 2017.

\bibitem{b10} C. Carrizo, M. Knapek, J. Horwath, D.D. Gonzalez, and P. Cornwell, ``Optical Inter-Satellite Link Terminals for Next Generation Satellite Constellations,'' in \emph{Proc. Society of Photo-Optical Instrumentation Engineers}, vol. 11272, article id. 1127203, 2020, pp. 1--11.

\bibitem{b10aa} D. Bhattacherjee and A. Singla, ``Network Topology Design at 27,000 km/hour,'' in \emph{Proc. 15th International Conference on Emerging Networking Experiments and Technologies}, Orlando, FL, USA, 2019, pp. 341--354.

\bibitem{b10a1} Y. Kaymak, R. Rojas-Cessa, J. Feng, N. Ansari, M. Zhou, and T. Zhang, ``A Survey on Acquisition, Tracking, and Pointing Mechanisms for Mobile Free-Space Optical Communications,'' \emph{IEEE Communications Surveys \& Tutorials}, vol. 20(2), pp. 1104–1123, Feb. 2018.

\bibitem{b10a2} T. Ahmmed, A. Alidadi, Z. Zhang, A.U. Chaudhry, and H. Yanikomeroglu, ``The Digital Divide in Canada and the Role of LEO Satellites in Bridging the Gap,'' \emph{IEEE Communications Magazine}, vol. 60(6), pp. 24--30, Jun. 2022. 

\bibitem{b10b} H. Hauschildt, C. Elia, A. Jones, H.L. Moeller, and J.M.P. Armengol,	``ESAs ScyLight Programme: Activities and Status of the High thRoughput Optical Network “HydRON”,'' in \emph{Proc. Society of Photo-Optical Instrumentation Engineers}, vol. 11180, article id. 111800G, 2019, pp. 174--181.
	
\bibitem{b10c} H. Hauschildt, C. Elia, H.L. Moeller, W. El-Dali, T. Navarro, M. Guta, S. Mezzasoma, and J. Perdigues, ``HydRON: High thRoughput Optical Network,'' in \emph{Proc. Society of Photo-Optical Instrumentation Engineers}, vol. 11272, article id. 112720B, 2020, pp. 57--67.

\bibitem{b10c1} C. Vasko, J. Perdigues, G. Acar, P. Arapoglou, W. El-Dali, S. Mezzasoma, M. Politano, Z. Sodnik, C. Elia, and H. Hauschildt, ``HydRON: Internet Backbone Beyond the Cloud(s),'' in \emph{Proc. Society of Photo-Optical Instrumentation Engineers}, vol. 11993, article id. 119930K, 2022, pp. 163--170.

\bibitem{b10d} H. Hauschildt, S. Mezzasoma, H.L. Moeller, M. Witting, and J. Herrmann, ``European Data Relay System Goes Global,'' in \emph {Proc. 2017 IEEE International Conference on Space Optical Systems and Applications}, Naha, Japan, 2017, pp. 15--18.

\bibitem{b10e} D. Calzolaio, F. Curreli, J. Duncan, A. Moorhouse, G. Perez, and S. Voegt, ``EDRS-C -- The Second Node of the European Data Relay System is in Orbit,'' \emph{Acta Astronautica}, vol. 177, pp. 537--544, 2020.

\bibitem{b12} M. Handley, ``Delay is Not an Option: Low Latency Routing in Space,'' in \emph{Proc. 17th ACM Workshop on Hot Topics in Networks}, Redmond, WA, USA, 2018, pp. 85--91.

\bibitem{b13} M. Handley, ``Using Ground Relays for Low-Latency Wide-Area Routing in Megaconstellations,'' in \emph{Proc. 18th ACM Workshop on Hot Topics in Networks}, New York, NY, USA, 2019, pp. 125--132.

\bibitem{b14} Y. Hauri, D. Bhattacherjee, M. Grossmann, and A. Singla, ``“Internet from Space” Without Inter-Satellite Links?,'' in \emph{Proc. 19th ACM Workshop on Hot Topics in Networks}, New York, NY, USA, 2020, pp. 205--211.

\bibitem{b15} S. Kassing, D. Bhattacherjee, A.B. Águas, J.E. Saethre, and A. Singla, ``Exploring the “Internet from Space” with Hypatia,'' in \emph{Proc. 2020 ACM Internet Measurement Conference}, New York, NY, USA, 2020, pp. 214--229.

\bibitem{b16a} A.U. Chaudhry and H. Yanikomeroglu, ``When to Crossover from Earth to Space for Lower Latency Data Communications?,'' \emph{IEEE Transactions on Aerospace and Electronic Systems (Early Access)}, doi:10.1109/TAES.2022.3156087.

\bibitem{b17} AGI, ``Systems Tool Kit (STK),'' [Online]. Available: \url{https://www.agi.com/products/stk}, Accessed: July 11, 2022.

\bibitem{b17a} NetworkX Developers, ``NetworkX – Network Analysis in Python,'' [Online]. Available: \url{https://networkx.org/}, Accessed: July 11, 2022.

\bibitem{b18} E.W. Dijkstra, ``A Note on Two Problems in Connexion with Graphs,'' \emph{Numerische Mathematik}, vol. 1, pp. 269--271, Dec. 1959.

\bibitem{b18a1} M. Barbehenn, ``A Note on the Complexity of Dijkstra's Algorithm for Graphs with Weighted Vertices,'' \emph{IEEE Transactions on Computers}, vol. 47(2), p. 263, Feb. 1998.

\bibitem{b18a2} J.S.B. Mitchell, ``Geometric Shortest Paths and Network Optimization,'' in \emph{Handbook of Computational Geometry}, Amsterdam: Elsevier Science, 2000.

\bibitem{ref18b} R.J. MacKay and R.W. Oldford, ``Scientific Method, Statistical Method and the Speed of Light,'' \emph{Statistical Science}, vol. 15(3), pp. 254--278, Aug. 2000. 

\bibitem{b19} R. Hermenier, C. Kissling, and A. Donner, ``A Delay Model for Satellite Constellation Networks with Inter-Satellite Links,'' in \emph{Proc. 2009 International Workshop on Satellite and Space Communications}, Siena, Italy, 2009, pp. 3--7.

\bibitem{b19a} R. Stanojevic, R.N. Shorten, and C.M. Kellett, ``Adaptive Tuning of Drop-Tail Buffers for Reducing Queueing Delays,'' \emph{IEEE Communications Letters}, vol. 10(7), pp. 570–572, Jul. 2006.

%\vspace{-1cm}

\end{thebibliography}
\end{document}